\documentclass[aps, prb, twocolumn, superscriptaddress, nofootinbib, nobibnotes, showkeys]{revtex4-1}%
\usepackage{epsfig}
\usepackage{amsmath}
\usepackage{amssymb}
\usepackage{xcolor}
\usepackage{xspace}
\usepackage[sort&compress]{natbib}
\usepackage{cancel}
\usepackage[normalem]{ulem}
\usepackage{hyperref}
\usepackage{graphicx}
\hypersetup{
	bookmarks=true,         
	unicode=true,          
	pdftitle={Role of hydrodynamic flows in chemically driven droplet division},    
	pdfauthor={Rabea Seyboldt and Frank Jülicher},     
	pdfkeywords={hydrodynamic flows, phase separation, nonequilibrium shape dynamics}, 
}

\newcommand{\Eqref}[1]{\mbox{Eq.\hspace{0.25em}\eqref{#1}}}
\newcommand{\Eqsref}[1]{\mbox{Eqs.\hspace{0.25em}\eqref{#1}}}
\newcommand{\figref}[1]{\mbox{Fig.\hspace{0.25em}\ref{#1}}}

\newcommand{\vect}{\boldsymbol}

\begin{document}

\title{Role of hydrodynamic flows in chemically driven droplet division}

\author{Rabea Seyboldt}
\email[Email: ]{seyboldt@pks.mpg.de}
\affiliation{%
	Max Planck Institute for the Physics of Complex Systems,
	01187 Dresden, Germany}
\author{Frank J\"ulicher}
\email[Email: ]{julicher@pks.mpg.de}
\affiliation{%
	Max Planck Institute for the Physics of Complex Systems,
	01187 Dresden, Germany}
\affiliation{%
	Center for Systems Biology Dresden,
	01307 Dresden, Germany}

\begin{abstract}
We study the hydrodynamics and shape changes of chemically active droplets. In non-spherical droplets, surface tension generates hydrodynamic flows that drive liquid droplets into a spherical shape. Here we show that spherical droplets that are maintained away from thermodynamic equilibrium by chemical reactions may not remain spherical but can undergo a shape instability which can lead to spontaneous droplet division. In this case chemical activity acts against surface tension and tension-induced hydrodynamic flows. By combining low Reynolds-number hydrodynamics with phase separation dynamics and chemical reaction kinetics we  determine stability diagrams  of spherical droplets as a function of dimensionless viscosity and reaction parameters. We determine concentration and flow fields inside and outside the droplets during shape changes and division. Our work shows that hydrodynamic flows tends to stabilize spherical shapes but that droplet division occurs for sufficiently strong chemical driving, sufficiently large droplet viscosity or sufficiently small surface tension. Active droplets could provide simple models for prebiotic protocells that are able to proliferate. Our work captures the key hydrodynamics of droplet division that could be observable in chemically active colloidal droplets.
\end{abstract}

\keywords{hydrodynamic flows, phase separation, nonequilibrium shape dynamics}
\maketitle

\addcontentsline{toc}{section}{Main text}
Living cells are compartmentalized in order to organize their biochemistry in
space.  Many cellular
compartments do not possess membranes and are formed by
the assembly of proteins and RNA in compact condensates
\cite{Brangwynne2009,Brangwynne2011,Hyman2012,Li2012,Weber2012,Brangwynne2013,Toretsky2014,Zwicker2014,Elbaum2015,Molliex2015,Patel2015,Feric2016,Saha2016,Banani2017,Sokolova2013,Lin2015}. Such condensates often have liquid like properties and resemble droplets that form by phase separation of a complex mixture \cite{Brangwynne2009,Patel2015,Feric2016,Saha2016}. Indeed protein droplets are observed to form in vitro by phase separation in physiological buffer \cite{Sokolova2013,Aumiller2015,Frankel2016,Saha2016,Nakashima2017}. 
Such droplets can organize chemical reactions in space, and the droplet dynamics can in turn be influenced by the reactions, as has been shown both in theory \cite{Glotzer1994,Puri1994,Christensen1996,Carati1997,Patashinski2012,Giomi2014,Zwicker2014,Zwicker2015} and experiments \cite{Crosby2012,Sokolova2013,Aumiller2015,Tang2015,Frankel2016,Saha2016,Nakashima2017}. 
The ubiquitous nature of RNA-protein condensates in a large variety of different cells and organisms suggests that the physical chemistry of macromolecular phase separation represents an evolutionary old mechanism for the compartmentalization
of chemistry and that droplet formation could have played a key role at the
origins of life and the emergence of prebiotic protocells \cite{Oparin1924,Oparin1952,Fox1976,Hanczyc2004,Browne2010,Koga2011,Murtas2013,Sokolova2013,Hyman2014,Li2014,Tang2014,ZwickerSeyboldt2017,Frankel2016,Lach2016}.

A minimal model of a protocell consists of a droplet that turns over by a chemical 
reaction and is constantly supplied with droplet material by diffusion from the outside \cite{ZwickerSeyboldt2017}. 
In such a scenario droplets are maintained away from thermodynamic
equilibrium and can reach a non-equilibrium 
steady state with a radius that is set by reaction parameters \cite{Zwicker2015}. 
An interesting possibility is that  the spherical shape of active droplets becomes unstable
and droplets spontaneously divide in two smaller daughters drops, 
providing a physical mechanism for the division of prebiotic cells \cite{ZwickerSeyboldt2017}. 
Such droplet dynamics is a hydrodynamic problem because surface tension in non-spherical
droplets drives hydrodynamic flows that redistribute material and deform the droplet shape 
\cite{Rayleigh1892,Chandrasekhar2013,Constantin1993,Paulsen2014}.

\begin{figure} [tb]
	\centering
	\includegraphics[width=0.8\columnwidth]{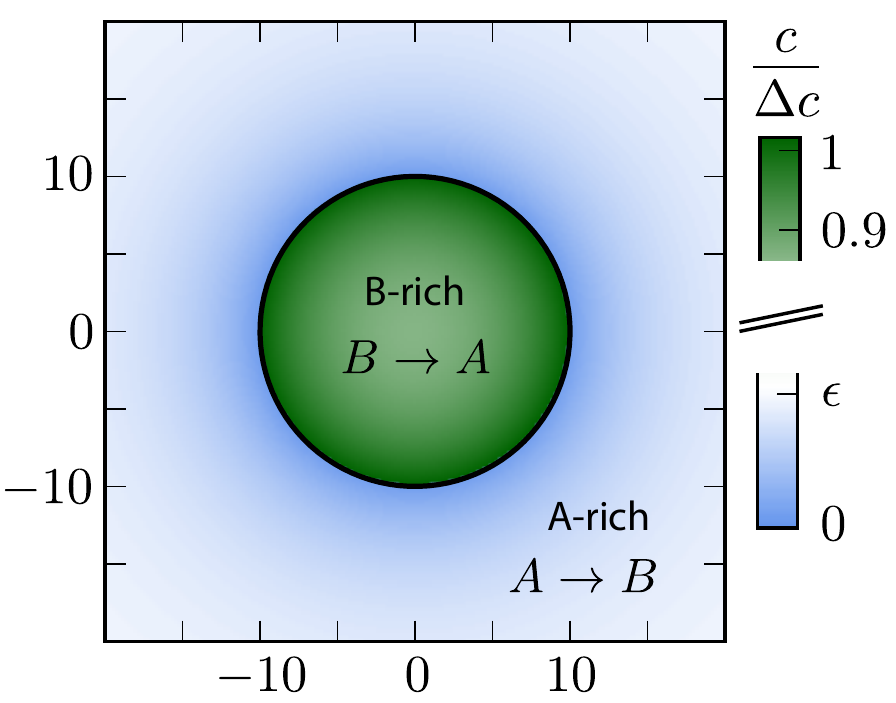}
	\caption{
		Chemically active droplet described by an effective droplet model. 
		Shown is the concentration field $c$ (blue and green color) of a stationary droplet (interface in black). Chemical reactions $B \to A$ create a sink of droplet material B in the droplet, and reactions $A \to B$ create a supersaturation $\epsilon$ of droplet material in the A-rich phase outside. This creates concentration gradients of B, which drive diffusion fluxes of droplet material, 
		while A flows in the opposite direction. The stationary droplet size results from the balance of the fluxes across the interface. 
		(Parameters: $\epsilon=0.176$, $A = 10^{-2}$, $\eta_+/\eta_-=1$, $k_+/k_-=1$, $\nu_-/(k_-\Delta c) = 1$, $D_+/D_- = 1$, $\beta_- = \beta_+$, $c^{(0)}_+=0$) 
		\label{fig:model}}
\end{figure}

Here we develop a hydrodynamic theory of the dynamics of chemically active 
droplets. We show that chemical reactions in active droplets
can perform work against surface tension and flows, 
giving rise to a shape instability that can result in droplet division.
We investigate the conditions for which droplets divide and determine
hydrodynamic flow fields of dividing droplets.

We  consider an incompressible liquid consisting of droplet material B and solvent component A which
can phase separate.
The local composition is characterized by the concentration field $c({\bf x})$ of component B. 
Volume preserving chemical reactions can transform component A into component B and back, 
 $A \rightleftharpoons B$. 
For simplicity, we first discuss an effective droplet model. A single droplet characterized by high concentration $c$ of
component B coexists with the surrounding fluid that mainly consists of A and contains 
B at low concentration, see \figref{fig:model}. Both phases are separated by a sharp interface.
The concentration of B satisfies a balance equation, where the chemical reaction provide a source or sink term 
$s_\pm(c)$,
\begin{eqnarray}
\partial_t c + \nabla \cdot \vect j = s_\pm (c)
\label{eq:c} \\
\vect j = - D_\pm \nabla c + \vect v c  \;.
\label{eq:c2}
\end{eqnarray}
Here, the indices $+$ and $-$ refer to quantities outside and inside the droplet, respectively. 
The flux $\vect j$ consists of advection by the fluid velocity $\vect v$ and a diffusion flux,
where $D_\pm$ denotes the diffusion constant of the droplet material in the two phases.

The chemical reaction is described by the concentration-dependent 
rate $s_\pm (c)$ which in general is a nonlinear function of $c$. 
 For simplicity, we linearize the chemical reaction rates
 in the vicinity of reference concentrations $c_\pm^{(0)}$ in each phase:
\begin{equation}
s_\pm (c) \simeq -k_\pm (c - c_\pm^{(0)}) \pm \nu_\pm \; ,
\label{eq:s}
\end{equation}
where $c_\pm^{(0)}$ are the equilibrium concentrations that coexist at equilibrium across a flat
interface. We have defined the reaction rate $\nu_\pm=s(c_\pm^{(0)})$ and the reaction constants $k_\pm=ds(c_\pm^{(0)})/dc$.
We focus on the case of positive coefficients $k_\pm>0$ and $\nu_\pm>0$, where B is produced outside the droplet, and degraded inside, see \figref{fig:model}. 

The hydrodynamic flow velocity ${\vect v}$ obeys Stokes equation of an incompressible fluid, 
\begin{equation}
\eta_\pm \nabla^2 \vect v = \nabla p \;,
\label{eq:stokes0}
\end{equation}
which accounts for momentum conservation $\partial_\alpha \sigma_{\alpha \beta}=0$, where
the stress tensor is given by $\sigma_{\alpha \beta} = \eta_\pm (\partial_\alpha v_\beta + \partial_\beta v_\alpha) - p \delta_{\alpha \beta}$. Here  $\eta_\pm$ denotes the fluid shear viscosities inside and outside of the droplet. 
The pressure $p$ plays the role of a Lagrange multiplier to impose the
constraint  $\nabla \cdot \vect v = 0$. 

The bulk equations (\ref{eq:c}-\ref{eq:stokes0}) are connected by boundary conditions at the droplet interface 
which we parameterize in spherical coordinates by the radial interface position $R(\theta,\phi)$ as a function of the polar and azimuthal angles  $\theta$ and $\phi$. The stress boundary conditions read
\begin{eqnarray}
\sigma_{nn}^+(R) - \sigma_{nn}^-(R) &=& 2 \gamma H(R) \\
\sigma_{nt}^+(R) - \sigma_{nt}^-(R) &=& 0  \;,
\label{eq:bc}
\end{eqnarray}
where 
$H(R)$ is the local mean curvature of the interface and  $\gamma$ is the droplet surface tension. The stresses at the interface on the inner and outer side of the droplet are denoted by $\sigma^\pm_{\alpha\beta}(R)$. 
The tensor indices $n$ and $t$ refer to tensor components normal and tangential to the interface, respectively.
The normal and tangential tensor components are defined as $\sigma^\pm_{nn}=
n_\alpha \sigma^\pm_{\alpha\beta} n_\beta$
and $\sigma^\pm_{nt}=n_\alpha \sigma^\pm_{\alpha\beta} t_\beta$, where $n_\alpha$ is a unit vector normal to the surface
and $t_\alpha$ is a unit vector tangent to the surface. Eq (\ref{eq:bc}) is valid for all tangent vectors and summation
over repeated indices is implied.
Using no-slip boundary conditions, the velocity field is continuous at the interface,
\begin{equation}
\vect v^+(R) = \vect v^-(R)  \;.
\label{eq:vn}
\end{equation}
The concentration field $c$ is discontinuous across the interface with values given by
\begin{eqnarray}
c_-(R) &=& c_-^{(0)} + \beta_- \gamma H(R) \label{eq:bcc} \\
c_+(R) &=& c_+^{(0)} + \beta_+ \gamma H(R) 
\label{eq:bcc2}
\end{eqnarray}
which are set by the physics of phase coexistence and a local equilibrium assumption.
The coefficients $\beta_\pm$ describe the effects of the Laplace pressure on the equilibrium concentrations at phase coexistence.
In the presence of fluxes at the interface, 
the interface moves in normal direction.  
The radial growth velocity is 
\begin{eqnarray}
\frac{dR}{dt} &=& \frac{\vect n}{\vect n \cdot \vect e_r} \cdot \frac{ \vect j^-(R)-\vect j^+(R)}{c_-(R) - c_+(R)}
\; ,
\label{eq:Rdot}
\end{eqnarray}
where $\vect n$ is a unit vector normal to the surface and $\vect e_r$ is a unit vector in radial direction. Eq. (\ref{eq:Rdot})
captures both convection of the interface by flows and droplet growth and shrinkage by addition or removal of material.

\begin{figure*}[tp]
\centering\includegraphics[width=\textwidth]{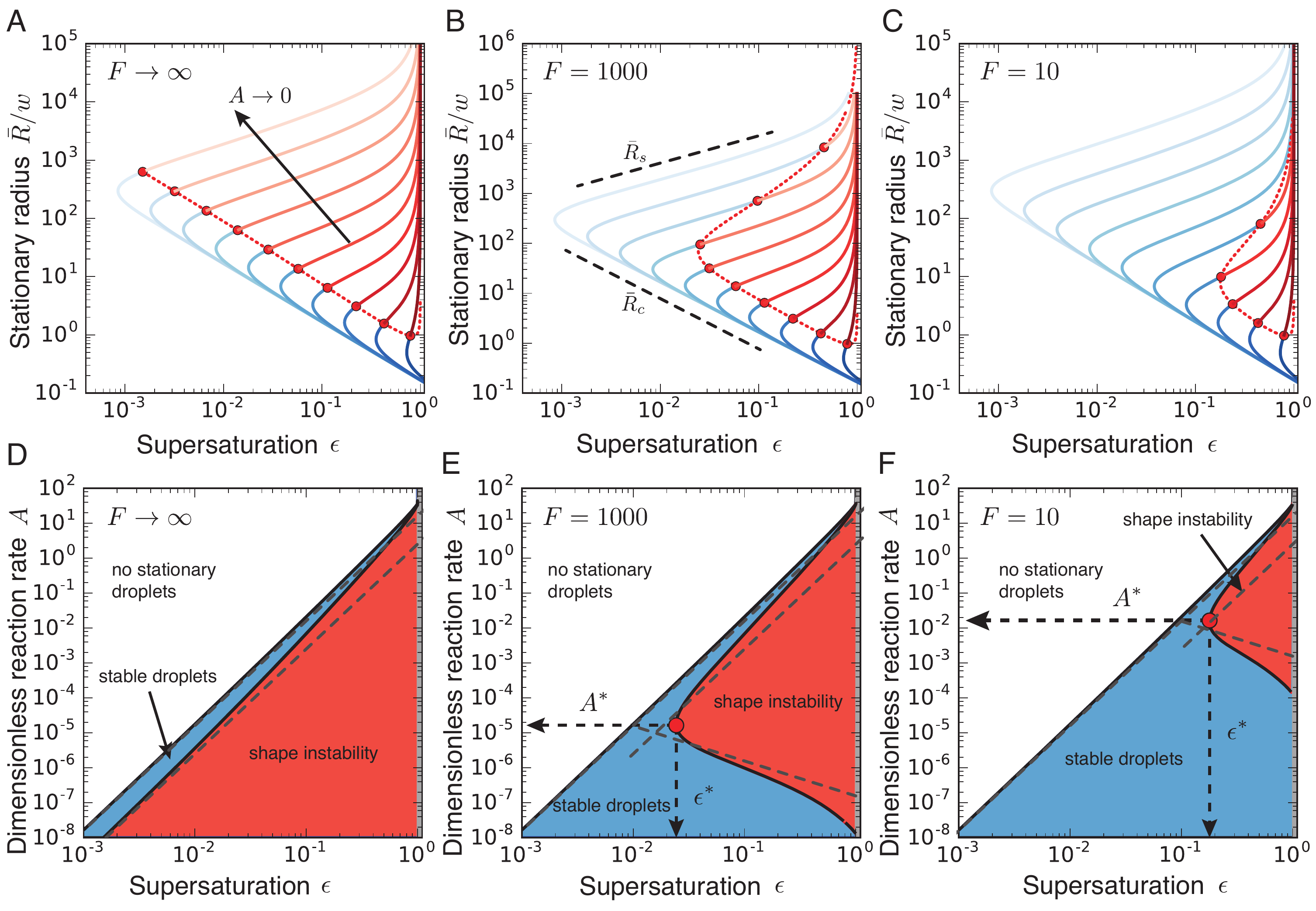}
\caption{\small
	 Stationary radii and onset of shape instability. \\
	A--C: Stationary radius as a function of supersaturation for different reaction amplitudes  $A=10^{-8}, 10^{-7}, \dots, 10^{1}$. The  stationary radii (lines) are independent of the dimensionless viscosity $F=w \eta_-/(\gamma \tau)$, while the onset of instability (red dots, connected by dotted red line) for the different curves varies in the three figures, which show dimensionless viscosities $F=\infty, 1000, 10$ (left to right). The blue line colors mark stable, the red ones unstable stationary radii with respect to the elongational $l=2$ mode. In panel B the scaling behavior of the nucleation radius $\bar R_c$ and the stationary radius $\bar R_s$ are indicated. 
	\\
	D--F: Stability diagram of stationary droplets of size $\bar R_s$, as a function of reaction amplitude $A$ and supersaturation $\epsilon$ for different dimensionless viscosities $F=\infty, 1000, 10$ (left to right). 
	For small supersaturation and large reaction amplitudes, no stationary radius exists (white). For large supersaturation, the stationary radius diverges (gray). In the region between these regimes, the stationary solution can be stable (blue) or unstable (red) with respect to shape perturbations of the $l=2$ mode. For decreasing $F$, the stable regime grows, and the minimal supersaturation $\epsilon^*$ at which an instability can be found increases, as well as the corresponding reaction amplitude $A^*$.  
	The scaling relations (dashed lines) for the regime of stable droplets and the onset of instability are indicated, with prefactors according to \ref{a:scal}. 
	(Parameters: $\eta_+/\eta_-=1$, $k_+/k_-=1$, $\nu_-/(k_- \Delta c) = 1$, $D_+/D_- = 1$, $\beta_- = \beta_+$, $c^{(0)}_+=0$) 
	 \label{fig:radii}  \label{fig:stab_diag}
 }
\end{figure*}


We find nonequilibrium steady state solutions to Equations (1-10) with a  spherical droplet of 
stationary radius $\bar R$ and stationary concentration field $\bar c(r)$, 
where $r$ is the radial coordinate, see Appendix A. The
stationary pressure $\bar p$ exhibits a jump $2\gamma/\bar R$ across the interface and no 
hydrodynamic flows exist, $\bar{\vect v}=0$.  An example for a stable non-equilibrium steady state with steady state concentration profile inside and outside the droplet of radius $\bar R$ is shown in \figref{fig:model}.
 
We first discuss the properties of these stationary states as a function of
external supersaturation $\epsilon= \nu_+/(k_+ \Delta c)$ and the dimensionless reaction rate $A= \nu_- \tau/\Delta c$ inside the droplet. The supersaturation is in our system generated by reactions outside the droplet and in steady state
corresponds to the concentration for which $s_+=0$. 
Here, $\Delta c = c_-^{(0)} - c_+^{(0)}$ and  we have introduced the time scale 
$\tau = w^2/D_+$, where $w = 6 \beta_+\gamma/\Delta c$ is a characteristic length scale. 
The stationary radii as a function of supersaturation $\epsilon$  are shown In \figref{fig:radii}A-C as solid lines for different values of  $A$.  
For values of $\epsilon$ smaller than a threshold value $\epsilon_0$,  no stationary radius exists. 
For values $\epsilon>\epsilon_0$ two steady state radii $\bar R_c$ and $\bar R_s$ exist, which 
become equal at $\epsilon_0$ where
they approach a value $\bar R_0$.  The smaller steady state radius 
$\bar R_c$ is a critical nucleation radius	 similar to the critical droplet radii found in passive systems. 
The larger radius denoted $\bar R_s$ stems from the interplay of phase separation and chemical reactions \cite{Zwicker2015,ZwickerSeyboldt2017}.	 
As the supersaturation reaches a value $\epsilon_{\infty}=\sqrt{(D_-k_-)/(D_+k_+)} \nu_-/(k_- \Delta c)$, the stationary radius $\bar R_s$ diverges.
	
We can find simple expressions for the stationary radii in the limit of small $A$ 
while keeping the ratios $\nu_-/(k_- \Delta c)$ and $k_+/k_-$ of reaction parameters fixed.
In this limit, the chemical reactions fluxes vanish as $s_\pm \propto A$ and the	
threshold value $\epsilon_0$ vanishes as $\epsilon_0 \propto A^{1/3}$. 
The critical nucleation radius then behaves as $\bar R_c \simeq w/(6 \epsilon)$ and the larger steady state
radius $\bar R_s \simeq w (3\epsilon A)^{1/2}$ where $\epsilon_0\ll \epsilon \ll \epsilon_{\infty}$, see 
\figref{fig:radii}B and \ref{a:scal}. 

The steady state solutions are independent on the fluid viscosity, however the droplet dynamics
is affected by hydrodynamic effects. We now investigate the role of hydrodynamic flows on chemically
driven shape
instabilities that can give rise to droplet division.
We perform a linear stability analysis at the stationary state given by $\bar X=(\bar c,\bar R, \bar p, \bar {\vect v} )$
for small perturbations $\delta X = (\delta c,\delta R,\delta p,\delta \vect v)$. 
The dynamics of these perturbations can be represented using eigenmodes
\begin{equation}
\delta X = \sum_{n,l,m} 
\epsilon_{nlm} X_{nlm} e^{\mu_{nlm} t} \;,
\end{equation}
with $X_{nlm}=(c_{nl} Y_{lm}, \bar R Y_{lm}, p_{l} Y_{lm}, \vect v_{lm})$, 
where  $Y_{lm}(\theta, \phi)$ are spherical harmonics with angular mode indices with 
$l=0,1,\dots$ and $m=-l,\dots,l$. The index $n=0,1,\dots$ denotes radial modes. 
The eigenmodes exhibit an exponential time dependence with a relaxation rate given by the eigenvalue $\mu_{nlm}$.
The mode amplitudes are denoted $\epsilon_{nlm}$.
The concentration modes are characterized by the radial functions $c_{nl}(r)$. The pressure modes
are described by $p_l(r)$ and the velocity modes $\vect v_{lm}(r,\theta,\varphi)$ 
can be expressed as 
\begin{equation}
\vect v_{lm} = v_{lm}^r \vect Y_{lm} + v_{lm}^{(1)} \vect \Psi_{lm} 
 +v_{lm}^{(2)} \vect \Phi_{lm} \;. 
\label{eq:v}
\end{equation}
where $\vect Y_{lm}(\theta, \varphi) = \vect e_r Y_{lm}$, $\vect \Psi_{lm}(\theta, \varphi) = r \nabla Y_{lm}$ and 
$\vect \Phi_{lm}(\theta, \varphi) = {\vect e}_r \times \vect \Psi_{lm}$ are vector spherical harmonics \cite{Barrera1985} and the radial functions
$v_{lm}^r(r)$, $v_{lm}^{(1)}(r)$ and $v_{lm}^{(2)}(r)$ characterize the velocity field. 
The radial functions can be obtained by solving the linearized dynamic equations using the corresponding boundary
conditions, see \ref{a:ana}.
The Stokes equation can be solved for a given shape perturbation independent of the concentration field so that the velocity field and pressure field is independent of the radial mode $n$. 
The radial part of the concentration field obeys a Helmholtz equation with an inhomogeneity that stems from
hydrodynamic flows. The homogeneous part is 
solved by modified spherical Bessel functions and the inhomogeneous solution can be found using Greens functions.
Using the dynamic equation for the shape changes of the droplet  \Eqref{eq:Rdot}, we obtain
an equation for the eigenvalue $\mu_{nlm}$,
\begin{equation}
\begin{split}
\mu_{nlm} = \frac{v_{l}^r(\bar R)}{\bar R}   +\frac{D_+}{\Delta c}\left (  {\bar c}''(\bar R_+) 
+ \frac{c_{nl}'(\bar R_+)}{\bar R}  \right )  \\
- \frac{D_-}{\Delta c} \left ( \bar c''(\bar R_-) 
+ \frac{c_{nl}'(\bar R_-)}{\bar R} \right )  \;.
\label{eq:mu}
\end{split}
\end{equation}
Here, the primes denote radial derivatives. 
Note that \Eqref{eq:mu} is an implicit equation for the eigenvalues $\mu_{nlm}$ because the radial concentration 
modes $c_{nl}(r)$ depend on $\mu_{nlm}$, see \ref{a:ana}. 
\Eqref{eq:mu} is independent of the index $m$, therefore the degeneracy of an eigenvalue $\mu_{nl}$ is at least $2l+1$. 
When all $\mu_{nl}$ are negative, the spherical shape is stable. The modes with $l=0$ correspond to changes in droplet size without flows.
They are always stable for $\bar R=\bar R_s$ and unstable for $\bar R=\bar R_c$. Thus droplet smaller than $\bar R_c$ will vanish and droplets larger will grow towards the size $\bar R_s$. Thus we consider the stability of $\bar R=\bar R_s$ in the following. 
The modes with $l=1$ do not involve shape deformations of the droplet and are thus not associated with flows. There always
exists a marginal mode with $\mu_{l=1}=0$ corresponding to overall translations where the droplet and all concentration fields are displaced and then stay in the new position. Here we consider shape instabilities for which a mode with $l>1$ becomes unstable. 
Because shape deformations induce flows, this instability depends on the 
dimensionless viscosity $F=w \eta_-/(\gamma \tau)$, as well as the ratio of viscosities in the two phases, $\eta_+/\eta_-$. 

If we increase the supersaturation $\epsilon$ while keeping the other parameters fixed,
 the steady state can become unstable with respect to the mode $l=2$ for a critical value $\epsilon=\epsilon_c$.
In \figref{fig:stab_diag}A-C, the onset of instability $\mu=0$ for the largest eigenvalue $\mu$ of the stationary radius is shown as a red dot, and unstable radii are indicated by red lines. Different lines correspond to different supersaturations, and the panels show different values of $F$. 
In \figref{fig:stab_diag}D-E, the corresponding stability diagrams of stationary droplets are shown as a function of the supersaturation and the reaction amplitude for different values of $F$. For large $A$ and small $\epsilon$, no stationary radius exists (white regions), so that any droplet would shrink and disappear. 
For large $\epsilon$, the stationary state diverges (gray regions). Spherical droplets are stable in the blue regions. Stationary spherical droplets are unstable inside the red region, the surrounding black line marks the shape instability with respect to the l=2 mode.  
The region where spherical droplets undergo a shape instability exists for $\epsilon\geq 
\epsilon^*$, which depends on $F$.  The value of $A$ for which the shape instability occurs at
$\epsilon=\epsilon^*$ is denoted $A^*$, see Fig 2E.

For small $A$,  the onset of instability can be describes by simple scaling behaviors. 
In the limit of small $A$ and for $\epsilon\ll \epsilon_{\infty}$, we find 
$\epsilon^*\propto F^{-1/2}$ and $A=A^*$ with $A^*\sim F^{-3/2}$ (compare Fig. 2E-F). 
For $A<A^*$, hydrodynamic flows govern the onset of instability which occurs at a value of $A$ which
behaves as $A\propto \epsilon^{-1} F^{-2} $. For $A>A^*$, hydrodynamic flows can be neglected 
as compared to diffusion fluxes and the onset of instability occurs for 
$A \propto \epsilon^{3}$. These two scaling regimes are indicated in
in \figref{fig:stab_diag}D-F by dashed lines.  
A derivation of these results including prefactors is given in \ref{a:scal}.

\begin{figure}[tb]
\centering\includegraphics[width=\columnwidth]{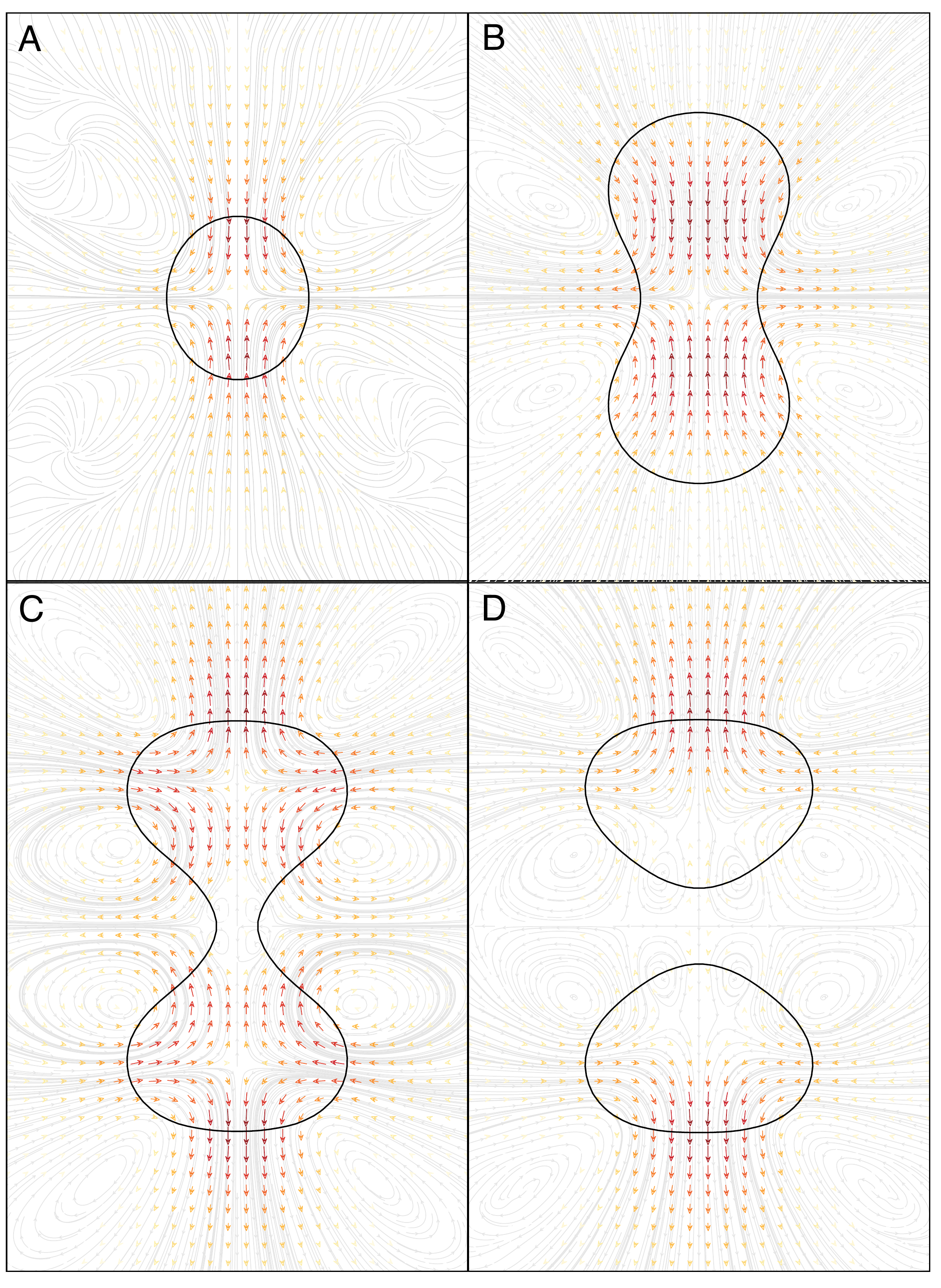}
\caption{ 
Numerical solution in 3d of an extended Cahn-Hilliard model with chemical reactions and hydrodynamic flows reveals that droplets can divide despite the presence of hydrodynamic flows. 
Panels A-D correspond to time points $t/\tau=100, 2100, 2700, 2800$, respectively, where $\tau=w^2/D$ is a diffusion time, with diffusion constant $D$ and interfacial width $w$.
The dynamic equations were solved numerically in a three-dimensional box. Shown are two-dimensional cross-sections of the droplet shape (black) together with streamlines (grey). 
Arrows (colored) indicate the direction and magnitude of the flow (normalized by respective maximal velocities $v_{max}\cdot w/D=0.0016$(in A), $0.0048$(B), $0.0034$(C) and $0.0047$(D) ).
(Parameters: $F=24$, $A = 8 \cdot 10^{-3}$, $\epsilon=0.2$, $\eta_-/\eta_+=1$, $c^{(0)}_+/\Delta c=0$, $k_+/k_-=1$, $\nu_-/(k_-\Delta c)=0.8$) 
 \label{fig:num}
 }
\end{figure}

We next address the question whether the shape instability found in the linear stability analysis
can indeed give rise to droplet divisions
in the presence of hydrodynamic flows in the nonlinear regime of the dynamics. 
We use a Cahn-Hilliard model \cite{Cahn1958}
for phase separation dynamics, extended to include chemical reactions and hydrodynamic flows, that can capture topological changes of the interface. 
We include chemical reactions via a source term linear in the concentration 
as well as advection by the hydrodynamic flow which is described by the incompressible Stokes
equation. Using a semi-spectral method \cite{Chen1998}, we obtain numerical solutions in a cubic box with no-flux boundary conditions, see \ref{a:num}. 

Starting from a weakly deformed spherical droplet, we find regimes where the droplet
disappears, where it relaxes to a stable spherical shape and where it undergoes
a shape instability, consistent with the linear stability analysis of the effective droplet model.
The transitions between these regimes occur for parameter values close to those predicted
by the linear stability analysis.
In the unstable regime, droplets typically divide. This shows that the droplet division
reported previously can also occurs  in the presence of hydrodynamic flows. 
\figref{fig:num} shows snapshots of the droplet shape together with corresponding hydrodynamic flow fields on the symmetry plane of a dividing droplet at different times. 
At early times when the droplet deformation is weak, the flow field is similar to 
the $l=2$ mode obtained from
the linear theory, \figref{fig:num} A. As the droplet elongates and its
waistline shrinks,  the flow field becomes more complex, see \figref{fig:num} B,C. 
The flow field shown in \figref{fig:num} C exhibits two additional vortex lines 
that form rings around the axis of rotational symmetry. Similarly, after division,
two further vortex rings occur, see \figref{fig:num} D.
Interestingly, for small deformations the hydrodynamic flow direction opposes the 
directions of interface motion at the main droplet axes, see \figref{fig:num} A,B. 
For larger deformations at later
times the flow switches its direction along the long droplet axis where it assists 
interface motion. At the waistline, the flow velocity becomes small, see \figref{fig:num} C.
After division, the flow field between the daughter droplets has very small magnitude, while strong
flows at the outer sides move the droplets apart \figref{fig:num} D. 

This example shows that division of active droplets can occur even if
hydrodynamic flows that oppose division are taken into account. Because flows
act in opposition to the initial deformation of the sphere, the linear stability
analysis already provides the key information of whether droplet division can
occur for a given value of dimensionless viscosity $F$,  
see \figref{fig:stab_diag}. 
This raises the question under what experimental conditions active droplets
would become unstable and division could be observed. 
Ignoring hydrodynamic flows, $F\to \infty$, it was shown that oil-water droplets and soft colloidal liquids or p-granules with sizes of a few micrometers could divide in the presence of chemical reactions \cite{ZwickerSeyboldt2017}. 
To address the influence of hydrodynamic flows, we
have to estimate the dimensionless viscosity $F=w\eta_-/(\gamma\tau)\simeq k_B T/(6\pi \gamma wa)$, where we have used $\tau=w^2/D$ and $D\simeq k_B T/(6\pi\eta a)$ with
molecular radius $a$. Thus, $F$  is
an equilibrium property of the phase separating fluid. For an oil-water system, we
estimate $F \approx 0.1 $, see \ref{a:para}. For soft colloidal liquids or p-granules, we estimate values between
$F \approx 10-10^{4} $. 
We can discuss these values using the stability diagrams in \figref{fig:stab_diag}D-F. Oil-water like droplets with $F \approx 0.1 $ are unlikely to divide, as the unstable region in the stability diagram is very narrow. For soft colloidal systems with $F \approx 10-10^{4} $, droplet division might be experimentally observable. 
We can estimate typical reaction rates required for division to occur based on the reaction rate $A^*$ for which the range of supersaturation is maximal. The value of $A^*$ corresponds to a reaction rate in the droplet of the order of $\nu_- = 10^{-4}mM/s$, see \ref{a:para}. A comparison with reported enzymatic reaction rates \cite{Stenesh2013} suggests that such values can be achieved in real systems.

We have shown that spontaneous division of chemically active droplets 
involves  mechanical work against surface tension as droplets deform.
Active droplets thus can transduce chemical energy  to mechanical work and
droplet division is therefore a mechano-chemical process.
The surface tension of the droplet creates pressure gradients as the droplet becomes non-spherical that lead to hydrodynamic flows. 
Because the flows generated act against the shape deformation,
droplets divide only for sufficiently large  viscosity or sufficiently small 
surface tension and sufficiently large  reaction rates.  
We show that the dependence of the onset of stability on parameters is captured for small reaction
fluxes by simple scaling relations.  
Our work shows that droplet division would
be suppressed in oil-water systems due to large surface tension and low viscosity. 
However it could be realized in soft colloidal systems for 
chemical reaction parameters that could be achieved experimentally. 
Furthermore flux-driven droplet divisions could be observable in 
biological systems, as both chemical reactions and phase-separating membrane-less 
organelles with low surface tensions can be found within cells.

\onecolumngrid
\clearpage
\appendix

\section{Effective droplet model with hydrodynamic flows}
\label{a:ana}

\subsection{Stationary state of a spherical active droplet}
Here, we discuss stationary solutions to 
equations (1-10) in the main text with spherical symmetry and without
hydrodynamic flows $\bar {\vect v} = 0$, where the bar indicates a steady state value. 
In this case, the pressure is constant both inside and
outside the droplet, with a pressure difference due to Laplace pressure between the inside and outside of the droplets, 
\begin{eqnarray}
	\bar p_- &=& \bar p_+ + \frac{2\gamma}{\bar R} \;.
\end{eqnarray}

The steady state concentration profiles in the presence of chemical reactions
are given by\cite{ZwickerSeyboldt2017}, 
\begin{eqnarray}
	\bar c_+(r) &=& + \frac{\nu_+}{k_+} + c_+^{(0)}  + A_+ k_0(r/l_+)\\
	\bar c_-(r) &=& - \frac{\nu_-}{k_-} + c_-^{(0)}  + A_- i_0(r/l_-) \;,
	\label{eq:cstat}
\end{eqnarray}
where $i_0(x)=2\sinh(x)/x$ and $k_0(x)=e^{-x}/x$ denote modified spherical Bessel functions of order zero of the first and second kind, respectively. 
The characteristic length scales $l_\pm = (D_\pm/k_\pm)^{1/2}$ are set by 
reaction rate constants and diffusion coefficients. 
The parameters $A_\pm$ are determined by the boundary condition at the droplet interface, Eq. (8-9) in the main text, 
\begin{eqnarray}
	A_+ &=& \left( \frac{ \gamma \beta_+}{\bar R} - \frac{\nu_+}{k_+} \right) \frac{1}{k_0(\bar R/l_+)} \\
	A_- &=&  \left( \frac{ \gamma \beta_-}{\bar R} + \frac{\nu_-}{k_-} \right) \frac{1}{i_0(\bar R/l_+)} \;.
\end{eqnarray}
Stationarity of the droplet radius $\bar R$ implies
\begin{equation}
	D_+ \bar c_+'(\bar R) = D_- \bar c_-'(\bar R) \; ,
	\label{eq:jbar}
\end{equation}
see Eq. (10) in the main text. 
Note that this equation typically has zero, one or two solutions for a given set of parameters. 

\subsection{Linearized dynamics}
We introduce small perturbations to the spherically symmetric stationary state, with $p=\bar p+\delta p$, $\vect v=  \delta \vect v$,
$c=\bar c+\delta c$ and $R=\bar R+\delta R$
and write the dynamics of these perturbations to linear order.
The linearized dynamics reads
\begin{eqnarray}
	\vect \nabla \delta p &=& \eta_\pm \Delta \delta \vect v \label{eq:perturbStokes}\\
	\vect \nabla \cdot \delta \vect v &=& 0 \label{eq:perturbIncomp} \\
	\partial_t \delta c &=& - \delta v_r \bar c' + D_\pm \nabla^2 \delta c - k_\pm \delta c \\
	\partial_t \delta R &=& \delta v_r(\bar R) + 
	\frac{1}{\Delta c} 
	\left[ D_+ \bar c_+ '' (\bar R) - D_- \bar c_-'' (\bar R) \right] \delta R  \nonumber\\
	&&  + \frac{1}{\Delta c} \left[ D_+ \partial_r \delta c_+(\bar R) - D_- \partial_r \delta c_-(\bar R) \right] \;.
	\label{eq:dRdot}
\end{eqnarray}
Here $\delta v_r$ denotes the radial part of the hydrodynamic velocity. 
With $\delta c_-$ and $\delta c_+$ we denote perturbations of the concentration field inside and outside the droplet. The same notation holds for the other fields. In this linear analysis, 
boundary conditions apply at the stationary radius $\bar R$,
\begin{eqnarray}
	\delta c_\pm(\bar R) &=& \beta_\pm \gamma \delta H - \bar c_\pm '(\bar R) \delta R \;,
	\label{eq:bc_phi}
\end{eqnarray}
with perturbation of the curvature 
$\delta H = H(R) - H(\bar R)$. 

The linearized dynamics can be decomposed in spherical harmonics, see Eq (11) in the main text.
The curvature perturbation then takes the form
\begin{equation}
	\delta H = \sum_{nlm}  \frac{h_l}{\bar R} \epsilon_{nlm} Y_{lm}\;,
	\label{eq:H}
\end{equation}
with $h_l = (l^2 + l - 2)/2$. 

\subsection{Hydrodynamic eigenmodes of the linearized dynamics }
We can expand the hydrodynamic eigenmodes using 
a basis of vector spherical harmonics, see Eq. (12) in the main text. 
The velocity boundary conditions Eq. (7) in the main text for the mode amplitudes
read
\begin{eqnarray}
	0 &=& v^{r+}_{lm}(\bar R) - v^{r-}_{lm}(\bar R) \\
	0 &=& v^{(1)+}_{lm}(\bar R) - v^{(1)-}_{lm}(\bar R) \\
	0 &=& v^{(2)+}_{lm}(\bar R) -  v^{(2)-}_{lm}(\bar R) \; . 
\end{eqnarray}
The stress boundary conditions 
(see Eq. (5-6) in the main text) at the interface read 
\begin{eqnarray}
	0 &=& 2 \eta_+  (v_{lm}^{r+})'(\bar R) - p_{lm}^+ (\bar R) 
	- 2 \eta_-  (v_{lm}^{r-})'(\bar R) + p_{lm}^- (\bar R) - 2 \gamma \epsilon_{lm}\frac{h_l}{\bar R}\\
	0 &=& \eta_+ \left[ (v_{lm}^{(1)+})' (\bar R) + \frac{v_{lm}^{r+}(\bar R)}{\bar R} - \frac{v_{lm}^{(1)+} (\bar R) }{\bar R}     \right]  \\
	&&- \eta_- \left[ (v_{lm}^{(1)-})' (\bar R) + \frac{v_{lm}^{r-}(\bar R)}{\bar R} - \frac{v_{lm}^{(1)-} (\bar R) }{\bar R}     \right] \\
	0  &=& \eta_+  \left[ (v_{lm}^{(2)+})' (\bar R) - \frac{v_{lm}^{(2)+} (\bar R) }{\bar R} \right] 
	- \eta_-  \left[(v_{lm}^{(2)-})' (\bar R) -\frac{v_{lm}^{(2)-} (\bar R) }{\bar R}   \right] \;.
\end{eqnarray}

We solve the radial profiles of the modes with a polynomial ansatz and exclude
functions that diverge for $r\to 0$ or $r\to\infty$ inside and outside the droplet, respectively.  
The pressure is then given by
\begin{eqnarray}
	p_{lm}^{-}(r) &=& \gamma f_{A} \left(\frac{r}{R}\right)^{l+1}  \\
	p_{lm}^{+}(r) &=& - \gamma f_{B} \left(\frac{r}{R}\right)^{-l}  \;.
\end{eqnarray}
For the hydrodynamic flow velocity we obtain
\begin{eqnarray}
	v_{lm}^{r-}(r) &=& \frac{\gamma}{\eta_-} \left[ f_{C1} \left(\frac{r}{R}\right)^{l+1} -  f_{C3} \left(\frac{r}{R}\right)^{l-1} \right] \\
	v_{lm}^{(1)-}(r) &=& \frac{\gamma}{\eta_-}  \left[  \frac{l+3}{l(l+1)}f_{C1} \left(\frac{r}{R}\right)^{l+1} - \frac{l+1}{l(l+1)} f_{C3} \left(\frac{r}{R}\right)^{l-1} \right] \\
	v_{lm}^{(2)-}(r) &=& 0
\end{eqnarray}
and 
\begin{eqnarray}
	v_{lm}^{r+}(r) &=& \frac{\gamma}{\eta_-} \left[ - f_{C2} \left(\frac{r}{R}\right)^{-l} + f_{C4} \left(\frac{r}{R}\right)^{-l-2} \right] \\
	v_{lm}^{(1)+}(r) &=& \frac{\gamma}{\eta_-} \left[ \frac{l-2}{l(l+1)} f_{C2} \left(\frac{r}{R}\right)^{-l} - \frac{1}{l+1} f_{C4} \left(\frac{r}{R}\right)^{-l-2} \right] \\
	v_{lm}^{(2)+}(r) &=& 0 \;.
\end{eqnarray}
Here, we have defined 
\begin{eqnarray}
	f_A &=& \frac{\left(l - 1\right) \left(l + 1\right) \left(l + 2\right) \left(2 l + 3\right)}
	{ \Delta  \left(2 l^{2} + 4 l\right) + \left(2 l^{2} + 4 l + 3\right)}
	\\
	f_B &=& \frac{ l \left(l - 1\right) \left(l + 2\right) \left(2 l - 1\right)}
	{\left(2 l^{2} + 1\right) + \left(2 l^{2} - 2\right)/\Delta}
	\\
	f_{C1}  &=& \frac{1}{2}\frac{ l \left(l - 1\right) \left(l + 1\right) \left(l + 2\right)}
	{ \Delta  \left(2 l^{2} + 4 l\right) + \left(2 l^{2} + 4 l + 3\right)} 
	\\
	f_{C2} &=& \frac{1}{2} \frac{ l \left(l - 1\right) \left(l + 1\right) \left(l + 2\right)}
	{ \Delta \left(2 l^{2} + 1\right) +  \left(2 l^{2} - 2\right)}
	\\
	f_{C3} &=& \frac{1}{2}\frac{ l \left(l - 1\right) \left(l + 1\right) \left(l + 2\right) 
		\left(\Delta \left(2 l^{2} + 4 l + 3\right) +  \left(2 l^{2} + 4 l\right)\right)}
	{\left(\Delta \left(2 l^{2} + 1\right) +  \left(2 l^{2} - 2\right)\right) 
		\left(\Delta \left(2 l^{2} + 4 l\right) + \left(2 l^{2} + 4 l + 3\right)\right)}
	\\
	f_{C4} &=& 
	\frac{1}{2} \frac{ l \left(l - 1\right) \left(l + 1\right) \left(l + 2\right) \left(\Delta \left(2 l^{2} - 2\right) +  \left(2 l^{2} + 1\right)\right)}
	{\left(\Delta \left(2 l^{2} + 1\right) + \left(2 l^{2} - 2\right)\right)
		\left(\Delta \left(2 l^{2} + 4 l\right) + \left(2 l^{2} + 4 l + 3\right)\right)} \; ,
	\label{eq:fC}
\end{eqnarray}
where $\Delta = \eta_+/\eta_-$ denotes the ratio of the viscosities inside and outside the droplet.

\subsection{Concentration eigenmodes}
The equation for the radial part of the concentration eigenmode is
\begin{equation}
	\frac{1}{D_\pm} v_{l}^r (r) \bar c'(r) = \left[ \frac{1}{r^2} \frac{d}{dr} r^2 \frac{d}{dr} - \lambda_{nl}^{\pm 2} - \frac{l(l+1)}{r^2} \right] c_{nl}(r)
	\label{eq:phi}
\end{equation}
with 
\begin{equation}
	\lambda_{nl}^{\pm 2} = (k_\pm + \mu_{nl})/D_\pm \;.
	\label{eq:lambda}
\end{equation}
The boundary conditions at $\bar R$ are
\begin{equation}
	c_{nl}(\bar R_\pm) = \gamma \beta_\pm \frac{h_l}{\bar R} - \bar R\; \bar c'(\bar R_\pm)  \;. 
	\label{eq:al}
\end{equation}
The left-hand side of \Eqref{eq:phi} constitutes an inhomogeneity 
\begin{equation}
	f_{l}^\pm(r) = -\frac{1}{D_\pm} v_{l}^r (r) \bar c' (r) \;. 
	\label{eq:fl}
\end{equation}
The solution $c_{nl}^\pm(r)$ of the inhomogeneous equation \eqref{eq:phi}
that satisfies the boundary condition \Eqref{eq:al}
can be constructed from a particular solution $c_{nl, p}^\pm(r)$ 
of the inhomogeneous equation
to which solutions $c_{nl, h}^\pm(r)$ of the homogeneous equation with $f_l^\pm=0$
are added to satisfy
the boundary conditions, \Eqref{eq:al}. This can be expressed as  
\begin{eqnarray}
	c_{nl}^-(r) &=& \alpha_{nl}^- c_{nl, h}^-(r) + c_{nl, p}^-(r) \\
	c_{nl}^+(r) &=& \alpha_{nl}^+ c_{nl, h}^+(r) + c_{nl, p}^+(r) \;,
	\label{eq:deltac}
\end{eqnarray}
where the coefficients $\alpha_\pm$ read
\begin{equation}
	\alpha_{nl}^\pm = \frac{a_l^\pm -c_{nl,p}^\pm (\bar R) }{c_{nl,h}^\pm(\bar R)} \;,
\end{equation}
with  $a_{l}^\pm=c_{nl}^\pm (\bar R)$.

We are especially interested in the case of unstable modes with $\mu_{nl} > 0$. Therefore we focus on the solution of equation \eqref{eq:phi} for $\lambda_{nl}^{\pm 2} > 0$ and $k_\pm > 0$. 
In this case, the homogeneous equation with $f_l^\pm=0$ is a modified Helmholtz equation which is
solved by modified spherical Bessel functions, $c_{nl, h}^-(r) = i_l(\lambda_{nl}^- r)$ and $c_{nl, h}^+(r) = k_l(\lambda_{nl}^+ r)$, where $i_l$ and $k_l$ denote the modified spherical Bessel functions of first and second order, respectively. 
The particular solution of the inhomogeneous equation can be obtained by 
a Green's function approach,  
\begin{eqnarray}
	c_{l,p}^-(r) &= &
	\lambda_{nl}^- k_l(\lambda_{nl}^- r) \int_0^{r} \left[ i_l(\lambda_{nl}^- r_2) f_{l}^-(r_2) r_2^2\right] dr_2 \\
	&& + \lambda_{nl}^- i_l(\lambda_{nl}^- r) \int_{r}^{\bar R} \left[ k_l(\lambda_{nl}^- r_2) f_{l}^-(r_2) r_2^2\right] dr_2
	\nonumber \\
	c_{l,p}^+(r) &= &
	\lambda_{nl}^+ k_l(\lambda_{nl}^+ r) \int_{\bar R}^{r} \left[ i_l(\lambda_{nl}^+ r_2) f_{l}^+(r_2) r_2^2\right] dr_2 \\
	&& + \lambda_{nl}^+ i_l(\lambda_{nl}^+ r) \int_{r}^{\infty} \left[ k_l(\lambda_{nl}^+ r_2) f_{l}^+(r_2) r_2^2 \right] dr_2 \nonumber \;,
\end{eqnarray}
with the radial part of the inhomogeneity $f_{l}^\pm(r)$ given by \Eqref{eq:fl}. 
The explicit calculation of these functions has to be handled with care, since the  
functions $k_l$ and $i_l$ have divergences
for large and small arguments $r$ 
that cancel in the final result 
but can still lead to numerical difficulties when evaluated directly.

The derivative of the concentration profile at $\bar R$ can be expressed as 
\begin{eqnarray}
	c_{nl}'(\bar R_-) &= \frac{a_l^-}{\bar R} g_{l,i}(\lambda_{nl}^- \bar R)  
	+ \frac{c_{l,p}^- (\bar R)}{\bar R} \cdot \left[ g_{l,k}(\lambda_{nl}^- \bar R) - g_{l,i}(\lambda_{nl}^- \bar R)\right]  
	\label{eq:dphi1} \\
	c_{nl}'(\bar R_+) &= \frac{a_l^+}{\bar R} g_{l,k}(\lambda_{nl}^+\bar R)  
	+ \frac{c_{l,p}^+ (\bar R)}{\bar R} \cdot \left[  g_{l,i}(\lambda_{nl}^+\bar R) - g_{l,k}(\lambda_{nl}^+\bar R) \right] \;,
	\label{eq:dphi2}
\end{eqnarray}
with 
\begin{eqnarray}
	g_{l,i} (x) &= \frac{x i_l'(x)}{i_l(x)} \\
	g_{l,k} (x) &= \frac{x k_l'(x)}{k_l(x)} \;.
	\label{eq:g}
\end{eqnarray}
Using the equation for the shape perturbations \eqref{eq:dRdot}, and using Eqns \eqref{eq:dphi1} and \eqref{eq:dphi2}, we obtain
Eq. (13) in the main text.  This equation determines the eigenvalue $\mu_{nlm}$ of
the hydrodynamic modes.

\subsection{Scaling relations in the limit of small reaction fluxes}
\label{a:scal}

In the limit of small chemical reaction fluxes $s_\pm$ we obtain simple scaling expressions for stationary radii 
and their shape instability conditions. 
Here we present the method and discuss the results. 

\subsubsection{Stationary radius}
Here we discuss the stationary radius in the limit of small chemical reaction amplitude $A=\nu_-\tau/\Delta c$
while keeping the ratios $\nu_-/(k_- \Delta c)$ and $k_+/k_-$ of reaction parameters fixed. 
This corresponds to the curves $\bar R(\epsilon)$ shown in \figref{fig:radii}A for different values of $A$. We can identify two regimes in the figure. The first is the region of small $\epsilon$, $\epsilon \sim \epsilon_0$, which corresponds to the minimum of $\epsilon(\bar R)$. The second is the region of $\epsilon_\infty$ where the stationary radius diverges. For $A\to 0$, we see that $\epsilon_0$ goes to zero while $\epsilon_{\infty}$ stays constant, and both are connected by a straight line that indicates scaling behavior of $\bar R=\bar R_s$. This increasing separation between $\epsilon_0$ and $\epsilon_{\infty}$ (and the corresponding stationary radii) in the limit of small $A$ means that we can analyze the behavior of the stationary radius in these two regimes separately. 
For this we consider Equations (A.2) and (A.3) for the concentration field and (A.6) for the stationary radius. We can rewrite (A.6) to obtain an expression relating the supersaturation to the stationary radius, 
\begin{equation}
\epsilon = \frac{\beta_+ \gamma}{\Delta c \bar R} + \left( \frac{\beta_- \gamma}{\Delta c \bar R} + \frac{\nu_-}{k_-\Delta c } \right) \frac{D_-}{D_+} 
\frac{\frac{\bar R}{l_-} \coth \frac{\bar R}{l_-} - 1}{1 + \frac{\bar R}{l_+}} \;.
\label{eq:epsscal}
\end{equation}
In this limit of small $A$, the characteristic length-scales of the concentration field become large with $l_\pm \propto A^{-1/2}$. To find scaling regimes in equation (\eqref{eq:epsscal}), we change variables in \Eqref{eq:epsscal} 
from $(A,\bar R)$ to $(A,\hat R)$ 
with $\hat R=\bar R A^a /w$, where $a$ is an exponent. For $a=1/3$ we find the behavior of $\epsilon(R)$ close to $\epsilon_0$ and $\bar R_0$, 
\begin{equation}
\hat \epsilon  =  \frac{1}{6} \hat R^{-1} + \frac{1}{3} \hat R^{2} + O(A^{1/6})
\label{eq:Rscal}
\end{equation}
where $\hat \epsilon= \epsilon A^{-1/3}$ becomes independent of $A$ for small $A$. 
This function describes the supersaturation as a function of radius around the threshold value $\epsilon_0$. Due to the inverted presentation $\epsilon(\bar R)$ instead of $\bar R(\epsilon)$ the function captures both the nucleation radius $\bar R_c$ and the larger radius $\bar R_s$. 
The threshold value $\epsilon_0$ can be obtained from \Eqref{eq:Rscal} by minimizing 
$\hat \epsilon$ for fixed $A$ as $\partial \hat\epsilon/\partial \hat R=0$. It behave as
\begin{equation}
\epsilon_0 =  4^{-2/3} A^{1/3}+O(A^{1/2}) \quad .
\end{equation}
For large and small $\hat R$, Eq (\ref{eq:Rscal}) describes the steady radii $\bar R_s$ and $\bar R_c$,
respectively, for which $\epsilon \geq \epsilon_0$.
For large $\epsilon$, the critical radius obeys
\begin{equation}
\bar R_c \simeq  \frac{w}{6 \epsilon} \;,
\label{eq:Rnucl}
\end{equation}
while the larger stationary radius is 
\begin{equation}
\bar R_s \simeq w(3\epsilon A)^{1/2} \;. 
\label{eq:Rbar}
\end{equation}
In \figref{fig:radii}B, the scaling behaviors given by
\Eqref{eq:Rbar} and \Eqref{eq:Rnucl} are indicated by dashed lines.  
At $\epsilon=\epsilon_0$ both radii meet at $\bar R=\bar R_0$, where
\begin{equation}
\bar R_0= w(4 A)^{-1/3} +O(A^{-1/2})\;.
\label{eq:eps0}
\end{equation}
For  $a=1/2$, $\bar R/l_\pm$ becomes independent of $A$ and 
\begin{equation}
\epsilon = \frac{D_-}{D_+} \frac{\nu_-}{k_-\Delta c} 
\frac{\frac{\bar R}{l_-} \coth \left( \frac{\bar R}{l_-} \right) - 1}{1 + \frac{\bar R}{l_+}} + O(A^{1/2})
\end{equation}
For $\bar R/l_\pm\ll 1$, the stationary radius obeys \Eqref{eq:Rbar} and is thus the larger stationary radius $\bar R_s$. 
For $\bar R/ l_\pm\gg 1$, we obtain the divergence of $\bar R_s$ as $\epsilon$ approaches $\epsilon_{\infty}$ with 
\begin{equation}
\epsilon_{\infty} =\sqrt{\frac{D_-k_-}{D_+k_+}} \frac{\nu_-}{k_-\Delta c} \;.
\end{equation}

\subsubsection{Shape instability}

	We now discuss scaling relations for the onset of instability in the $(A,\hat \epsilon)$ 
	plane in the limit of small $A$, which give the trends shown as dashed lines in \figref{fig:radii}D-F.  
	We use the scaling of the stationary radius $\bar R=\bar R_s$ close to $\epsilon_0$ with $\hat R=\bar R A^{1/3} /w$, $\hat\epsilon=\epsilon A^{-1/3}$ and 
	$\hat l_\pm = l_\pm A^{1/2}$ in 
	\Eqref{eq:mu} to obtain
	\begin{equation}
	\hat \mu_{nlm} =  - \frac{d_l}{\hat R} \frac{A^{-2/3}}{F}  + \frac{2}{3}(l-1) - \frac{D_+}{D_-} \frac{ (l-1) g_l}{\hat R^3}  + O(A^{1/6})
	\label{eq:muscal}
	\end{equation}
	where $\hat \mu_{nlm} = \mu_{nlm} \tau/A$ and $\hat R$ is related to $\hat\epsilon$ by (\ref{eq:Rscal}).
	Here, $d_l=f_{C3}-f_{C1}$, where $f_{C1}$ and $f_{C3}$ are defined in \Eqref{eq:fC} and 
	\begin{equation}
	g_l =\frac{ h_l(l+1) + \frac{D_-}{D_+} \frac{\beta_-}{\beta_+}h_l l}{l-1}
	\end{equation}
	with $h_l = (l^2+l-2)/2$. 
	For large mode index $l$, 
		\begin{equation}
		d_l = \frac{l}{2(\eta_+/\eta_-+1)} + O(1/l)\;. 
		\end{equation}
	We now consider conditions for which $\mu_{nlm}=0$ for small $A$ and the mode $(n,l,m)$ becomes unstable. 
	Using \eqref{eq:Rscal} in \eqref{eq:muscal}, we find a relation between $\hat \epsilon$ and $\hat R$ at the onset of instability $\mu_{nlm}=0$, 
	\begin{equation}
	\hat \epsilon = \frac{d_l}{2(l-1)} \frac{1}{\hat F} \hat R + \left( \frac{1}{6} + \frac{D_+}{D_-} \frac{1}{2} g_l \right) \hat R^{-1} + O(A^{1/6}) \;. 
	\label{eq:Rinstab}
	\end{equation}
	This curve captures the scaling behavior of the onset of instability for different parameters in the $\bar R-\epsilon$ plane, corresponding to the red dotted line in \figref{fig:radii}A-C. 
	
	We now focus on finding the scaling relations for the onset of stability of the stationary radius as function of $A$, $\epsilon$ and $F$, as shown in \figref{fig:radii}D-F. At this onset, both \eqref{eq:Rinstab} and \eqref{eq:Rscal} need to be satisfied. We use both equations to eliminate $\hat R$. 
	We find a crossover regime with relations $A^* \sim F^{-3/2}$ between the region where hydrodynamic flows are relevant ($A<A^*$) and where they can be neglected ($A>A^*$). 
	For $A>A^*$ we find  for $\mu_{nlm}=0$ as relation between $A$ and $\epsilon$
	\begin{equation}
	A \simeq 54 \frac{g_l}{ \left(1+\frac{1}{2} g_l\right)^3   } \epsilon^{3} \;. 
	\label{eq:instab1}
	\end{equation}
	For $A<A^*$ we find 
	\begin{equation}
	A \simeq \frac{1}{3} \left( \frac{2(l-1)}{d_l} \right)^2 \epsilon^{-1} F^{-2} \;.
	\label{eq:instab2}
	\end{equation}
	In \figref{fig:stab_diag}D-F, the dashed lines indicate these two scaling solutions in the limit $A\to 0$ and $F\to \infty$ for $l=2$, which we find to be the first mode to become unstable. We find that the general trends of the stability diagram is captured well, with small deviations from the full solution of \Eqref{eq:mu} for small $\epsilon$, and larger deviations in the regime close to $\epsilon_\infty$ where the scaling of the stationary radius $\bar R_s\propto A^{-1/3}$ breaks down.

\newpage
\section{Continuum model for active droplets with flows}
\label{a:num}

\subsection{Continuum model for active droplets}
\label{sec:continuous_model}

We study an extended Cahn-Hilliard equation with chemical reactions coupled to Stokes equation for hydrodynamic flows at low Reynolds numbers. 
We consider an incompressible fluid 
containing two components $A$ and $B$, with number concentration fields $c_A(\vect r, t)$ and $c=c_B(\vect r, t)$ that depend on position $\vect r$ and time $t$, and with 
molecular masses $m_A$ and $m_B$ and molecular volumes $v_A$ and $v_B$. 
We are interested in the case where component $A$ forms 
the background fluid and $B$ is a droplet material that forms droplets 
by phase separation. Additionally, chemical reactions convert the two components into each other, $A \rightleftharpoons B$. 
For simplicity, we consider mass and volume conserving chemical reactions with 
$m_A/v_A = m_B/v_B$, which encodes that volume is conserved in the reaction if mass is conserved. 
Together with incompressibility, this implies that the mass density $\rho = m_A c_A + m_B c_B$ is constant.
Therefore, we can describe the system by the concentration $c(\vect r, t)$ of the droplet material $B$ only. 

We use the following double-well free energy density \cite{Cahn1958}
\begin{equation}
	f(c) =
	\frac{b}{2 (\Delta c)^2}
	\Bigl(c- c^{(0)}_-\Bigr)^2
	\Bigl(c- c^{(0)}_+\Bigr)^2
	+ \frac{\kappa}{2} \bigl(\nabla c \bigr)^2
	\label{eqn:free_energy_density}
	\;,
\end{equation}
with $\Delta c = \bigl| c^{(0)}_- - c^{(0)}_+ \bigr|$.
Here, the positive parameter $b$ characterizes  
molecular interactions and entropic contributions.
This free energy describes the segregation of the fluid 
in two coexisting 
phases \cite{Desai2009}: one phase rich in droplet material with $c \approx c^{(0)}_-$ and a dilute phase with $c \approx c^{(0)}_+$. 
The coefficient $\kappa$ 
is related to surface tension and the interface width~\cite{Cahn1958}.

The state of the system is characterized by the free energy
\begin{equation}
	F[c] = \int d^3 r \; f(c) \;
	\label{eqn:free_energy}
	\;,
\end{equation}
where the integral is over the system volume. We work with an ensemble $T$, $\rho$, $c$ here, where $T$ denotes temperature and the system is considered isothermal. 
The chemical potential $\bar\mu = \delta F[c]/\delta c$, governs demixing and  can be split into local and nonlocal contributions, $\bar \mu = \bar \mu_{0}-\kappa \nabla^2 c$ with
\begin{equation}
	\bar\mu_{0} =  
	\frac{b}{(\Delta c)^2}
	\bigl(c- c^{(0)}_+\bigr)
	\bigl(c- c^{(0)}_-\bigr)
	\bigl(2c- c^{(0)}_- - c^{(0)}_+\bigr)
	\; .
	\label{eqn:chemical_potential}
\end{equation}
The dynamics of the concentration field is described by \cite{deGroot2013,Prost2015}
\begin{eqnarray}
	\partial_t c &= -\nabla \cdot \vect j + s(c)
	\label{eqn:cahn_hilliard} \\
	\vect j &=- M \nabla \bar \mu + \vect v c
	\label{eqn:cahn_hilliard2}
	\; .
\end{eqnarray}
Here, $M$ is a mobility coefficient of the droplet material and $\vect v$ is the hydrodynamic velocity. 
The source term $s(c)$ describes chemical reactions, for which we choose for simplicity a linear concentration dependence, 
\begin{equation}
	s(c) = \nu - k (c-c_+^{(0)} )  \;.
	\label{eqn:source}
\end{equation}
The reaction flux given in \Eqref{eqn:source}
does not obey detailed balance with respect to the free energy, and thus describes a situation where an external energy source
maintains the system away from equilibrium \cite{ZwickerSeyboldt2017}. 

The hydrodynamic velocity $\vect v$ can be calculated using momentum conservation, 
\begin{equation}
	\partial_t (\rho v_\alpha) = \partial_\beta \sigma_{\alpha\beta} \;,
	\label{eq:momcons}
\end{equation}
with momentum $\rho v_\alpha$ and stress tensor $\sigma_{\alpha\beta}$, where $\alpha$ and $\beta$ number cartesian coordinates $x, y, z$. 
We can write the stress tensor $\sigma_{\alpha\beta}$ as
\begin{equation}
	\sigma_{\alpha\beta} = -(\rho v_\alpha) v_\beta + \sigma_{\alpha\beta}^{eq} + \sigma_{\alpha\beta}^{d} \;,
	\label{eq:sigma}
\end{equation}
where the first term describes advection of the stress tensor, $ \sigma_{\alpha\beta}^{eq}$ and $\sigma_{\alpha\beta}^{d}$ denote the equilibrium and dissipative stress tensors. The equilibrium stress tensor is given by 
\begin{equation}
	\sigma_{\alpha\beta}^{eq} = -(\bar \mu c- f) \delta_{	\alpha\beta} - \frac{\partial f}{\partial (\partial_\alpha c)} \partial_\beta c - P_0 \delta_{\alpha \beta} \;.
	\label{eq:sigmaeq}
\end{equation}
Here, $\bar \mu c - f$ is the osmotic pressure of the droplet material, and $\delta_{\alpha\beta}$ denotes the Kronecker delta. Incompressibility is enforced by an additional partial pressure $P_0$. 
The deviatory stress tensor can be found as thermodynamic force related to momentum by writing the entropy production rate, 
\begin{equation}
	\sigma_{\alpha\beta}^{d} = 2 \eta \left(v_{\alpha\beta} - \frac{1}{3} v_{\gamma \gamma} \delta_{\alpha\beta} \right) + \eta' v_{\gamma\gamma} \delta_{\alpha\beta} \;,
	\label{eq:sigmad}
\end{equation}
where $\eta$ and $\eta'$ denote viscosities, and $v_{\alpha\beta}=(\partial_\alpha v_\beta + \partial_\beta v_\alpha)/2$ is the symmetric strain tensor. 

In the Stokes limit, the inertial terms are neglected, $D_t (\rho v_{\alpha}) = 0$, with advected derivative $D_t = \partial_t + v_\beta \partial_\beta$, leaving $0 = \partial_\beta (\sigma_{\alpha\beta}^{eq} + \sigma_{\alpha\beta}^{d})$. This yields \cite{Cates2012}
\begin{eqnarray}
	\eta \partial_\beta^2 v_{\alpha} &= 3 \bar \mu_{0} \partial_{\alpha} c - \kappa c \nabla^2 (\partial_\alpha c) + \partial_{\alpha}P_0\;. 
	\label{eq:stokesc}
\end{eqnarray}

\Eqsref{eqn:chemical_potential}--\eqref{eqn:source} and \Eqsref{eq:stokesc} and incompressibility $\partial_{\alpha} v_\alpha=0$ 
define the continuum model of active droplets.

\subsection{Numerical solution of the continuum model}
\label{a:numnum}

We numerically solve the dynamic equations of the continuum model of active droplets, \Eqsref{eqn:chemical_potential}--\eqref{eqn:source} and \Eqsref{eq:stokesc} with \Eqref{eq:reproj1} and incompressibility $\partial_{\alpha} v_\alpha=0$. 

For this we use a spectral method in a 3d rectangular box. 
This has the advantage that in a spectral decomposition, the spatial operators become simple multiplications with the wavenumber \cite{Chen1998}. However, our equations contain a number of nonlinear functions, which are easier to evaluate in real space. We therefore transform forward and back in each time step. 

To calculate the next timestep $t_i$ from the fields found in timestep $t_{i-1}$, we use a semi-implicit Runge-Kutta method \cite{Ascher1997} (method (2,3,3)) for the concentration field. 
This evaluates the gradient term in $\bar \mu$, \Eqref{eqn:chemical_potential}, implicitly, while evaluating the rest of $\bar \mu$ as well as the advection term of the fluxes, $\vect v c$, explicitly. This effectively means that the terms related to the interfacial profile are calculated implicitly, which allows for larger time steps as an explicit scheme. 

For the concentration field, we choose no-flux boundary conditions ($\partial_n c = 0$, where the derivative is in a direction normal to the simulation box), which leads to a decomposition in cosine functions in the spectral description. 
The Laplacian then is $-k^2$ for a mode with wave vector $\vect k$. 
The Stokes equation can also be solved using spectral methods. Here, no-flux conditions lead to $v_n = 0$.  Additionally we enforce incompressibility using a reprojection method.  
For this, the velocity field calculated by neglecting the partial pressure, $P_p=0$, can be split into two parts (Helmholtz decomposition), 
\begin{equation}
	\vect v = \vect v_\psi +\vect  v_\phi = \nabla \times \vect \psi - \nabla \phi
\end{equation}
with vector field $\vect \psi$ and scalar field $\phi$, and velocity parts $\vect v_\psi = \nabla \times \vect \psi$ and $\vect v_\phi = - \nabla \phi$. 
With this, we find 
\begin{equation}
	\nabla \cdot \vect v = \Delta \phi
	\label{eq:reproj1}
\end{equation}
and thus, using incompressibility, $\nabla \cdot \vect v = 0$, we can calculate $\phi$. We thus find the incompressible part of the velocity field
\begin{equation}
	\vect v_\psi = \vect v - \nabla \phi \;.
	\label{eq:reproj2}
\end{equation}
We can evaluate this in Fourier space using a spectral method. 
For a rectangular box aligned with the coordinate system, we thus find that each velocity component $v_\alpha$ is decomposed by sines in one direction and cosines in the other direction. Spatial derivatives convert a sine-description into cosines, and vice versa. 

We normalize concentration, length, time and energy by 
$\Delta c = c_-^{(0)} - c_+^{(0)}$, $w = 2(\kappa/b)^{1/2}$, $t_0 = w^2/D$ and $\hat e_0 = \kappa \hat w (\Delta c)^2/3 $, respectively, where the characteristic length scale is
$w = 2(\kappa/b)^{1/2}$.
The relevant dimensionless model 
parameters are 
$c_+^{(0)} / \Delta c$, $k t_0 $, 
and $\nu_- t_0/  \Delta c$. 
We choose 
$c_+^{(0)} / \Delta c = 0 $,  $k t_0 = 10^{-2}$, $\nu t_0 = 2 \cdot10^{-3}$ and $\eta \; \hat w^3/(t_0 \hat e_0) = 2$. 
Additionally, we use as box-length $L / \hat w = 100$ in all 3 dimensions, number of grid-points in one direction $N=128$ and simulation time $T/t_0=4 \cdot 10^3$. For the time step, we start with a timestep of $\Delta t/t_0 = 10^{-4}$, and double the timestep to a final step size of $\Delta t /t_0= 0.01$. 

We start with initial conditions $R = R_0 (1+\epsilon Y_{2,0})$, with $R_0/\hat w = 7$ and $\epsilon = 1$. 
The concentration field at positions $\vect{r}$ is initialized by the function
\begin{equation}
	c(\vect r) = \frac{c_+^{(0)} + c_-^{(0)}}{2} + \frac{c_+^{(0)} - c_-^{(0)}}{2} \tanh \frac{d(\vect r)}{w} \; . \label{initial} 
\end{equation}
where $d(\vect r)$ is the oriented distance of $\vect r$ 
to the nearest point on the ellipsoid. The value of
$d(\vect r)$ is negative for points inside the droplet and positive for points outside.

\newpage
\subsection{Effective droplet model as a limit of the continuum model}
We now discuss the relationship between the effective droplet model and the continuum model. 
To relate the two models, we first use the continuum model to derive jump conditions for the concentration in the effective droplet model in equilibrium. We then consider stress balance across this interface and derive stress boundary conditions in the effective droplet model. Finally we discuss the dynamical equations in the bulk and at the interface in non-equilibrium situations. 

\subsubsection{Derivation of jump conditions for equilibrium phase separation}
First we consider the phase separation in equilibrium without chemical reactions in the continuum model. 

In a one-dimensional system with a mean concentration $\bar c$ with $c^{(0)}_+ < \bar c < c^{(0)}_-$, the free energy of the system in \Eqref{eqn:free_energy} is minimized by 
the concentration profile 
\begin{equation}
c^*(x) = \frac{c^{(0)}_- + c^{(0)}_+}{2} + \frac{c^{(0)}_- - c^{(0)}_+}{2}\tanh\frac{x}{w} \;,
\label{eq:ctanh}
\end{equation}
where $w=2(\kappa/b)^{1/2}$ denotes the interfacial width and $x$ is the normal distance to the interface.  The concentration profile describes two phases of concentration $c^{(0)}_-$ and $c^{(0)}_+$ separated by a flat interface of width $w$.  
The surface tension can be defined as 
\begin{equation}
\gamma = \int_{-\infty}^{\infty} F[c^*(x)] - \frac{1}{2}(F[c_-^{(0)}]+ F[c_+^{(0)}])dx \;. 
\end{equation}
For the free energy \Eqref{eqn:free_energy} with the concentration profile \Eqref{eq:ctanh}, this can be written as $\gamma = \int_{-\infty}^{\infty} \kappa (\nabla c^*)^2 dx$ which yields
$\gamma = (\Delta c)^2/6 \; \sqrt{\kappa b}$  or  \cite{Safran1994}. 

This interfacial tension governs the concentration jump condition in the effective droplet model, which can be derived as follows. 
To describe a curved interface, we consider two homogeneous phases with concentrations $c_\pm$. 
For a finite volume $V_s$ with a droplet of size $V$ and area $A$ the concentrations $c_\pm$ can be found by minimizing the free energy $F=f(c_-) V + f(c_+)(V_s-V) + \gamma A$ with $\partial F/\partial c_- |_V =0$ and $\partial F/\partial V|_{c_-} = 0$,  where the concentration of both phases are related by $V_s\bar c = V c_-+(V_s-V)c_+$ where $\bar c$ denotes the average concentration in the system. 
Thus for two phases to be in equilibrium, their chemical potential $\bar \mu$ and osmotic pressure $\Pi=c \bar \mu-f$ need to obey
\begin{eqnarray}
0 &=& \bar \mu(c_-) - \bar \mu(c_+) \\
0 &=& \Pi(c_+) - \Pi(c_-)  - 2 \gamma H \;, \label{eq:Laplacepressure}
\end{eqnarray}
where $H$ the mean curvature of the droplet and $2\gamma H$ is the Laplace pressure. These equations determine the concentrations in the phases $c_\pm$ of coexisting phases \cite{Safran1994}. 

For small Laplace pressures, we can express the equilibrium concentrations $c_\pm$ of a curved interface by the concentrations of a flat interface $c_\pm^{(0)}$ plus a small perturbation, 
\begin{eqnarray}
c_- &=& c_-^{(0)} + \beta_- \gamma H \\
c_+ &=& c_+^{(0)} + \beta_+ \gamma H 
\label{eq:cbc}
\end{eqnarray}
where $\beta_\pm = 2/(f''(c_\pm^{(0)}) \Delta c)$. For the free energy \Eqref{eqn:free_energy}, we find $\beta_\pm = 2/(b\Delta c)$, which is related to the interfacial width as 
$w = 6 \gamma \beta_+/\Delta c$.

\subsubsection{Stress balance across the interface}
\label{a:stressbalance}
\begin{figure}[tb]
	\centering\includegraphics[width=0.7\textwidth]{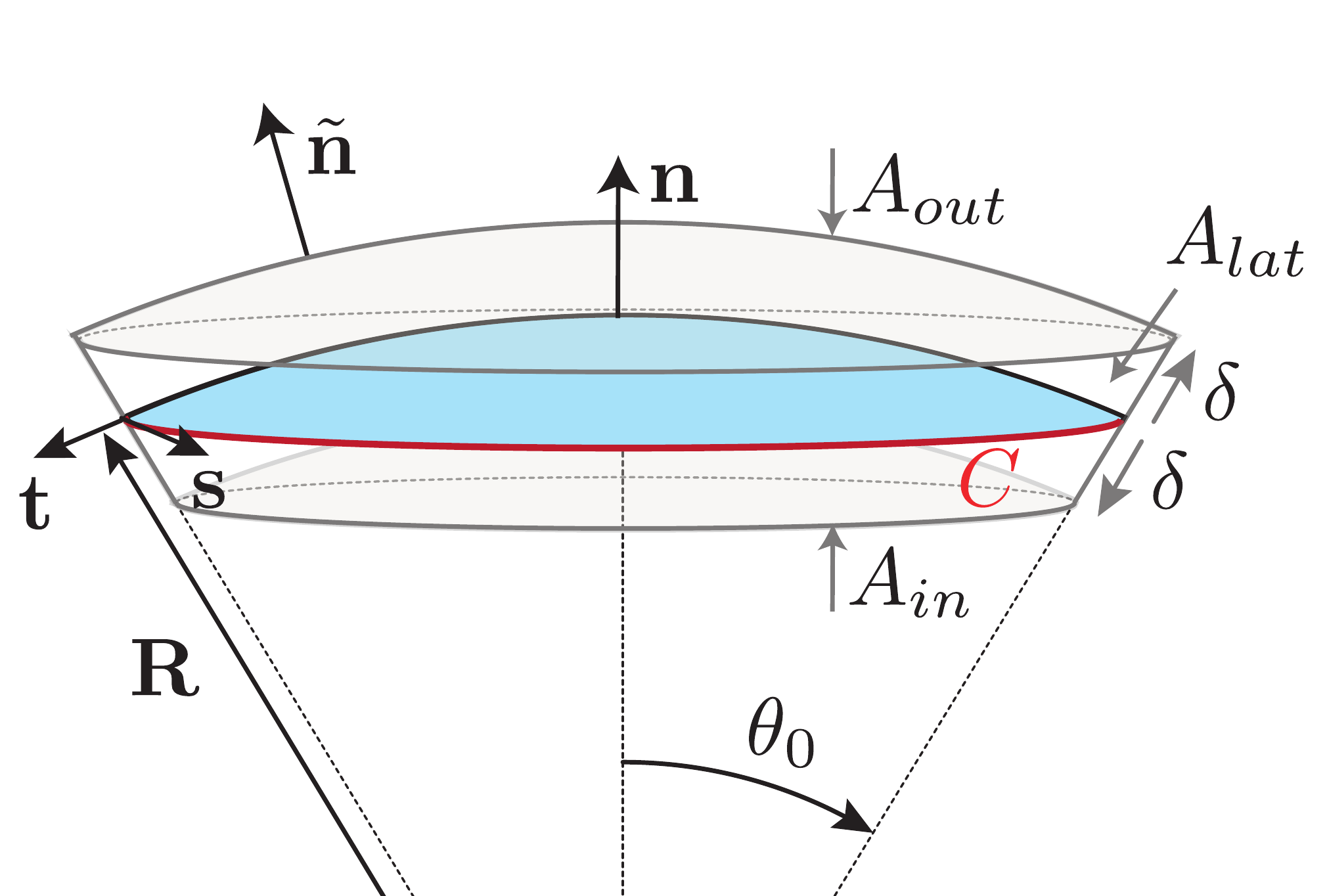}
	\caption{ 
		Geometry for the force balance. We consider a spherical cap of the droplet interface, with a box with constant distance $\delta$ to the interface inside and outside. The normal and tangential vectors $\vect n$, $\vect t$ and $\vect s$ of the interface are shown, as well as the normal vector $\vect {\tilde n}$ of the box. The origin of the spherical coordinate system is the center of the sphere that describes the interfacial curvature, with radius $\tilde R$, while $\theta_0$ gives the polar angle of the cap. 
		\label{fig:geo}
	}
\end{figure}

We now consider stress balance of the continuum model across the droplet interface to derive stress jump conditions at the interface in the effective droplet model. 
We discuss the mechanical equilibrium in a small volume across a curved interface with a local mean curvature $H$ corresponding to a (local) effective radius $\tilde R = 1/H$.  
We focus on the case where the interface is rotationally symmetric around the considered point $\vect R$, and where the curvature does not change along the interface. 
We use spherical coordinates, where the radial vector $\vect e_r$ is aligned with the (outward pointing ) normal vector $\vect n$ and the tangential vectors $\vect t$ and $\vect s$ are aligned with $\vect e_\theta$ and $\vect e_\phi$, respectively (with the vector directions for $\phi=0$ in the limit $\theta=0$). We consider a small box enclosing $\vect R$ where the outer and inner surfaces $A_{out}$ and $A_{in}$ have a constant distance of $\delta$ to the interface, and the lateral surface $A_{lat}$ is at a constant angle $\theta_0$ with respect to the symmetry axis. The geometry is shown in \figref{fig:geo}.

Now let us consider the balance of the stress tensor \Eqref{eq:momcons} across the box, taking into account the curved geometry. 
The stress balance $\partial_\beta \sigma_{\alpha \beta}$ can be written as 
\begin{equation}
0 = \oint dA \; \tilde{n}_\beta \sigma_{\alpha \beta}
\end{equation}
where $\alpha$ and $\beta$ are cartesian coordinates and $\tilde{n}$ the (local, outward pointing) normal vector of the box-surface. 
We can split this in three terms, 
\begin{equation}
0 = \int dA_{out}  \sigma_{\alpha n} - \int dA_{in}  \sigma_{\alpha n} + \int dA_{lat}  \sigma_{\alpha t} \;,
\label{eq:stressbox}
\end{equation}
where we used that the orientation of the normal vectors of the box coincides with the normal/tangential vector of the interface. 

On the inner and outer areas $A_{in}$ and $A_{out}$, the stress tensor presented in \Eqref{eq:sigma} with equilibrium stress tensor in \Eqref{eq:sigmaeq} reduces to the form of the effective droplet model given after \Eqref{eq:bc} in the main text, as the gradient terms are negligible for $\delta \gg w$. 
We now consider the limit of a sharp interface $w\to 0$ with finite surface tension $\gamma$, and consider the case of a small box of thickness $\delta$, which remains larger than the interfacial width. 
The components $\alpha=x,y$ of \Eqref{eq:stressbox} vanish by symmetry. For $\alpha=z$ we find
\begin{eqnarray}
0 =  \pi \tilde R^2 \sin^2 \theta_0 \; \sigma_{nn}^{+} - \pi \tilde R^2 \sin^2 \theta_0 \; \sigma_{nn}^{-}  - 2 \pi \tilde R \sin^2 \theta_0 \; \gamma  \;,
\end{eqnarray}
where $\sigma_{nn}^{\pm}$ are the stress tensor components of the effective model, \Eqref{eq:stokes0}, inside and outside the interface at $\vect R$. Integration over the lateral box surface $A_{lat}$ yields the last term, $\int dA_{lat}  \sigma_{\alpha t} \cong 2 \pi \tilde R \sin^2 \theta_0 \; \gamma$. 
We thus find that the mechanical equilibrium of a curved interface introduces a Laplace pressure $2\gamma H$, 
\begin{equation}
0 = \sigma_{nn}^{+} - \sigma_{nn}^{-}  - 2 \gamma H \;. 
\label{eq:Laplace}
\end{equation}
We therefore recover the stress jump conditions of the effective droplet model, \Eqref{eq:bc}. 
Additionally, \eqref{eq:Laplace} together with \eqref{eq:Laplacepressure} implies that the partial pressure needed to satisfy incompressibility is continuous across the interface, $P_0^+=P_0^-$.

\subsubsection{Dynamics of the effective droplet model}
We now consider the dynamics of a non-equilibrium system with a droplet. We show how the continuum model is related to the bulk equations and jump conditions of the effective droplet model. 
For this we consider a droplet with a interface that is thin compared to the dynamical length scales $l_\pm$, so that we can describe the interface by local equilibrium. In the bulk phases we focus on the case where deviations from the equilibrium concentrations are small.  

In the bulk phases, we expand the chemical potential \Eqref{eqn:chemical_potential} around the reference concentrations $c_\pm^{(0)}$. The gradient term $-\kappa \nabla^2 c$ in the chemical potential is important within the interface, but can be ignored in the bulk phases, where the length-scales on which the concentration field varies are much larger than the interfacial width. Thus we can describe the chemical potential by 
\begin{equation}
\bar \mu_\pm (c) \approx \left. \frac{d\bar \mu_0}{dc}\right|_{c_\pm^{(0)}} (c-c_\pm^{(0)}) \;,
\label{eq:linpot}
\end{equation}
which is $\bar \mu_\pm (c) \approx b (c-c_\pm^{(0)})$ for our specific free energy. 
With this simplification, Eqns. \eqref{eqn:cahn_hilliard} and \eqref{eqn:cahn_hilliard2} become the reaction-diffusion-convection equations \eqref{eq:c} and \eqref{eq:c2}  with diffusion constants $D_\pm = M \left.(d \bar \mu_{0}/dc)\right|_{c_\pm^{(0)}}$ or $D_\pm = M b$. 
Similarly we linearize the chemical reaction rate \Eqref{eqn:source} in both phases. As we already chose a linear rate for the continuum model, we only need to relate the parameters $k$ and $\nu$ with the constants $k_\pm$ and $\nu_\pm$ of the effective model, with $k_\pm = k$, $\nu_+ = \nu$ and $\nu_- = k \Delta c -\nu $.  
Inserting the linearized chemical potential \Eqref{eq:linpot} into the equilibrium stress tensor \eqref{eq:sigmaeq} we find that momentum conservation in the bulk phases is given by the Stokes equation \eqref{eq:stokes0} with viscosities $\eta_\pm = \eta$, where the pressure $p$ is determined by the incompressibility condition $\partial_{\alpha} v_\alpha = 0$. 

We consider the droplet interface to be in local equilibrium. We therefore obtain \Eqref{eq:bcc} for the jump of the concentration field in the effective model. 
The incompressibility condition $\partial_\alpha v_\alpha=0$ implies $ v^-_n (R) = v^+_n(R)$ at a sharp interface, and we consider an interface without slip length, so that $\vect v^- (R) = \vect v^+(R)$. 
We thus find \Eqref{eq:vn} of the effective model. 
The normal stress balance in \Eqref{eq:bc} is derived in \ref{a:stressbalance}. 

As a last point we need to find \Eqref{eq:Rdot} for the interface movement. 
We consider the concentration change in a box of width $\delta$ around the interface, see \figref{fig:geo}. We consider a box enclosing a point $\vect R$ on the interface at the time $t$ aligned with the normal and tangential directions of the interface at $\vect R$. The interface may move with normal movement $\partial_t \hat R(t)$, with $\hat R(t) = \vect R(t) \cdot \vect n$ and normal vector $\vect n$, while the box stays at a fixed position.
The total change of material in the volume is given by
\begin{equation}
\partial_t \int_V dV  \; c = - \int_A dA\; \vect {\tilde{n}} \cdot \vect j  + \int_V dV\; s(c) 
\label{eq:Vc}
\end{equation}
where $V$ denotes the volume and $A$ the area of the box. For small $w$ and finite $\delta$ the concentration field $c$ makes a jump from the surface $A_{in}$ to $A_{out}$ given by conditions \eqref{eq:bcc} and \eqref{eq:bcc2} at $\hat R$. Within each phase, we can express the field by the boundary values at the interface \Eqref{eq:cbc} and a linear expansion, 
\begin{equation}
c(\vect r,t) \simeq 
\begin{cases}
c_-(\vect R(t) ) + \nabla c_-(\vect r,t) \cdot (\vect r - \vect R(t) )   & \mbox{inside droplet}\\
c_+(\vect R(t) ) + \nabla c_+(\vect r,t) \cdot (\vect r - \vect R(t) )  & \mbox{outside droplet}
\end{cases}
\end{equation}
The chemical reaction is given in both phases by \Eqref{eqn:source}. 
For small $\delta$ and $\theta_0$, we find for the left-hand side of \Eqref{eq:Vc} that $\delta c$ vanishes to lowest order and 
\begin{equation}
\partial_t \int_V dV\; c = A_R (c_-(\vect R(t) ) - c_+(\vect R(t) )) \partial_t \hat R + O(\epsilon)+O(\theta_0)
\end{equation}
where $A_R$ is the area of the droplet interface enclosed by the box. For a spherical cap, $A_R = 2\pi (1-\cos \theta_0) \hat R^2$.  
We further find that the source term due to the chemical reaction scales with the volume of the box, and thus vanishes for a small box, $\int_V dV \; s(c)  = 0 + O(\epsilon)+O(\theta_0)$. 
The flux across the box can be expressed as 
\begin{equation}
 - \int_A dA\;  \vect {\tilde{n}} \cdot \vect j = A_R \vect n \cdot (\vect j_-(\vect R(t) ) - \vect j_+(\vect R(t) )) + O(\epsilon)+O(\theta_0)
\end{equation}
where $\vect j_\pm(\vect R(t) ) $ denotes the flux at $\vect R$ inside/outside the droplet. 
We thus find the normal movement of the interface, 
\begin{equation}
\partial_t \hat R  = \vect n \cdot \frac{ \vect j_-(\vect R(t) ) - \vect j_+(\vect R(t) ) }{ c_-(\vect R(t) ) - c_+(\vect R(t)) }\;. 
\label{eq:Rdot2}
\end{equation}
In the main text we use spherical coordinates centered at the droplet center. For a spherical droplet, the normal and radial movement would thus be the same. For a deformed droplet, we need to consider the relation between the normal interface movement, $\hat R(t) = \vect R (t) \cdot \vect n$ and the radial movement $R(t) = \vect R (t) \cdot \vect e_r$. At fixed angles $\theta$ and $\phi$, the interface movement is given by $\partial_t  {\vect R} = \partial_t R \;\vect e_r$. 
Using $\partial_t \hat R= \partial_t \vect R(t) \cdot \vect n$, we find a relation between the radial and normal movement, $\partial_t  R=\partial_t {\hat R}/(\vect n\cdot \vect e_r)$. This relation, together with \Eqref{eq:Rdot2}, yields the interfacial movement \Eqref{eq:Rdot} presented in the main text. 

We thus recover all dynamical equations of the effective droplet model from the continuum model based on irreversible thermodynamics. 
Note that the specific choice of the free energy leads to specific relations between parameters of the effective model such as $D_+=D_-$. 
Our derivation shows the relation between both models in the case where the interface width $w$ is small compared to the droplet size, $R/w\gg 1$, and the chemical diffusion length, $l_\pm/w\gg1$. 
Additionally, we focused on the case where the concentrations in the phases are similar to the concentrations in equilibrium and have small concentration gradients. 
These conditions are not valid in all systems. Most importantly, the chemical reactions can drive concentrations far away from the equilibrium phase concentrations $c_\pm^{(0)}$. The resulting behaviors, such as the formation of new interfaces associated with instabilities of the spinodal decomposition regime,
are not captured in the effective droplet model.

\clearpage
\subsection{Comparison of the droplet dynamics in the continuum model and the effective model}
\begin{figure}[tb]
	\centering\includegraphics[width=0.7\textwidth]{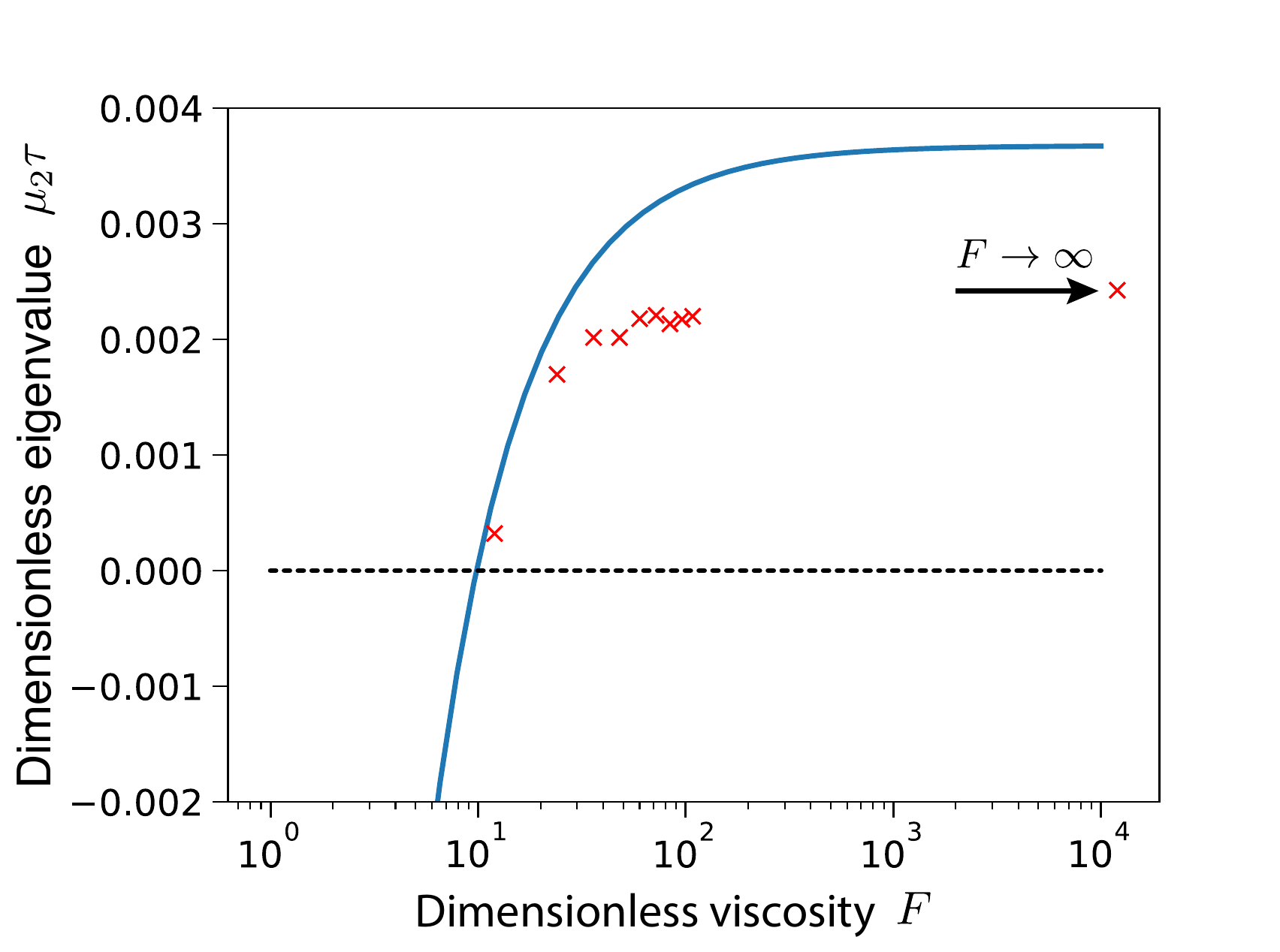}
	\caption{ 
		Growth of shape perturbations of the $l=2$ mode for different normalized viscosities $F=\eta w/(\gamma \tau)$ for the continuous model (red crosses) and effective model (blue curve). The last data point (with arrow) corresponds to $F\to\infty$. 
		 (Parameters: $A = 8 \cdot 10^{-3}$, $\epsilon=0.2$, $\eta_-/\eta_+=1$, $c^{(0)}_+/\Delta c=0$, $k_+/k_-=1$, $\nu_-/(k_-\Delta c)=0.8$)
		\label{fig:numS}
	}
\end{figure}

Here we compare the analytical predictions of the effective model for the instability with numerical calculations of the continuous model for different values of the renormalized viscosity $F$. 
For this we numerically solved the dynamic equations of the continuous model starting with a droplet with a small initial deformation of mode $l=2$. We fitted the dynamical behavior of the mode to an exponential function, with yields a numerical estimate for the eigenvalue $\mu_2$. 
In figure \ref{fig:numS} the resulting eigenvalues are shown, together with the eigenvalue of corresponding parameters of the effective model. 
We find that the value of F for which droplet shapes become unstable is very similar to the value predicted by the effective model. The eigenvalues are qualitatively similar to the ones of the effective model, despite working in an a parameter regime where the interfacial width and the differences of concentration within a phase cannot be considered very small, so that the models are not necessarily comparable. 

To generate the data in the figure, we initialized droplets with a small shape perturbation for different values of $F$. All parameters and initial conditions were chosen as described in \ref{a:numnum}. We found that for $F\geq 100$ droplets divide, while they are stable for $F \leq 1$. For $F=10$, the shape deformation was very slow, so that division was not seen in the time interval $T/\tau=4000$. For $10<F<100$, as well as $F=\infty$, we fitted radius and spherical harmonic deformation to the concentration field using \Eqref{initial}. For short times, the droplet radius changes as the concentration field and droplet size go towards the stationary values. After that, the shape deformation grows until the droplet deforms so strongly that the fitting fails. By hand we chose intermediate time windows for the simulations where the size was stationary and the shape deformation small. In these windows we fitted the deformation amplitude $\epsilon$ (compare \Eqref{initial}) with an exponential function, $A e^{\mu_2 t} + B$ with parameters $A$, $B$ and eigenvalue $\mu_2$ to the $l=2$ mode of the shape deformation. 

\clearpage
\section{Estimation of parameters}
\label{a:para}

Here we estimate the hydrodynamic parameter for two physical phase-separating systems to understand the importance of hydrodynamic flows on the droplet division in experimental systems. We discuss two cases, water-oil phase separation, and soft colloidal systems (such as protein-RNA phase-separation in cells). We have already estimated parameter values for both systems without the influence of hydrodynamic flows \cite{ZwickerSeyboldt2017}, where we found that droplet division should be possible for realistic values of chemical reaction rates in both systems, and that corresponding stationary radii would have sizes of a few micrometers. 
Here we estimate the value of the dimensionless viscosity $F$ for water-oil and soft colloidal systems, and compare them to the analytical phase diagrams presented in \figref{fig:stab_diag}. 

To calculate the hydrodynamic parameter $F$ for experimental systems, we need an estimation of the diffusion coefficient of the droplet material $D_+$ outside the droplet, 
of the interfacial width $w$ (which corresponds to length-scale $w$ in the paper \cite{ZwickerSeyboldt2017}), of the surface tension $\gamma$ and of the viscosity $\eta_-$ inside the droplet.
For water-oil systems, the interfacial width is of the order of $w \approx 1nm$ and the diffusion constant is $D_+ \approx 10^{-9} m^2/s$. 
We can estimate the  surface tension as $\gamma \approx 10^{-2} N/m$, and the viscosity  $\eta_- \approx10^{-3} (N\cdot s)/m^2$ \cite{Safran1994,Haynes2014}. 
With these values, we find $F \approx 0.1$. In this case droplet division is strongly 
suppressed, see Fig. 2 of the main text.
For soft colloidal systems, we estimate $w \approx 10nm$,
$D_+ \approx 10^{-10} m^2/s$ and $\gamma \approx 10^{-6} N/m$ \cite{Safran1994,Brangwynne2009}. The value of $F$ depends on the viscosity of the
droplet.  For values $\eta_- \approx10^{-3} (N\cdot s)/m^2$, $F \approx 10$, and 
for $\eta_- \approx 1-10 (N\cdot s)/m^2$, we have $F \approx 10^{4}$. In both cases
droplet division is possible, but more easy to achieve for larger $F$. 
We convert  $A^*$ to the reaction rate $\nu_-$ inside the droplet using the droplet concentration given in \cite{ZwickerSeyboldt2017}. 

We can use \Eqref{eq:instab1} and \eqref{eq:instab2} from the scaling analysis to estimate the instability of the concrete parameter examples discussed in \cite{ZwickerSeyboldt2017} under the influence of hydrodynamic flows. In these scaling equations, the ratios $\eta_+/\eta_-$ and $D_- \beta_- /(D_+\beta_+)$ enter the calculation of $A^*$ and $\epsilon^*$ but we find that they do not lead to relevant changes in the results. The scaling analysis thus yields results very similar to the estimation using Fig. 2.

\clearpage


\bibliographystyle{apsrev4-1}
\bibliography{papers}

\end{document}